\pdfoutput=1

\documentclass[universe,review,accept,moreauthors,pdftex,12pt,a4paper]{mdpi}
\firstpage{1} 
\articlenumber{x}
\doinum{10.3390/------}
\pubvolume{xx}
\pubyear{2016}
\copyrightyear{2016}
\externaleditor{Academic Editor: Lorenzo Iorio}
\history{Received: 5 January 2016; Accepted: 7 March 2016; Published: date}

\usepackage{graphicx,soul}

\usepackage{soul}

\Title{The Scales of Gravitational Lensing}

\Author{Francesco De Paolis $^{1,2,\dagger,}$*, Mos\`e Giordano $^{1,2,\dagger}$, Gabriele Ingrosso $^{1,2,\dagger}$, Luigi Manni $^{1,2,\dagger}$, \mbox{Achille Nucita $^{1,2,\dagger}$} and Francesco Strafella $^{1,2,\dagger}$}

\address{%
$^{1}$ \quad Department of Mathematics and Physics ``Ennio De Giorgi'', University
of Salento, {CP 193, I-73100 Lecce, Italy}; giordano@le.infn.it (M.G.); ingrosso@le.infn.it (G.I.); luigi.manni@le.infn.it (L.M.); nucita@le.infn.it~ (A.N.); strafella@le.infn.it (F.S.) \\

$^{2}$ \quad INFN (Istituto Nazionale di Fisica Nucleare), Sezione di Lecce CP-193, Italy\\
$^\dagger$ \quad These authors contributed equally to this work.}

\corres{Correspondence: francesco.depaolis@le.infn.it; Tel.: +39-0832-297493; Fax: +39-0832-297505}

\abstract{After exactly a century since the formulation of the general theory of relativity, the phenomenon of gravitational lensing
is still an extremely powerful method for investigating in astrophysics and cosmology. Indeed, it is adopted to study the distribution of the stellar component in the
Milky Way, to study dark matter and dark energy on very large scales and even to discover exoplanets. Moreover, thanks to technological developments, it will allow the measure of the physical parameters (mass, angular momentum and electric charge) of supermassive black holes in the center of ours and nearby galaxies.}



\begin{document}


\section{Introduction}

In 1911, while he was still involved in the development of the general theory of relativity (subsequently published in 1916), Einstein
made the first calculation of light deflection by the Sun~\cite{einstein1911}. He correctly understood that a massive body may act as a gravitational lens
deflecting light rays passing close to the body surface. However, his calculation, based on Newtonian mechanics, gave a deflection angle wrong by a factor of two.
{On 14 October 1913,} Einstein wrote to Hale, the renowned astronomer, inquiring whether it was possible to measure {a deflection angle of about $0.84''$ toward the Sun.}
The answer was negative, but Einstein did not give up, and when, in 1915, he made the calculation again using the general theory of relativity, he found the right value $\phi=2r_s/b$
(where $r_s=2GM/c^2$ is the Schwarzschild radius and $b$ is the light rays' impact parameter) that corresponds to an angle of about $1.75''$ in the case of the Sun.
That result was resoundingly confirmed during the Solar eclipse of 1919 \cite{dyson1919}.

In 1924, Chwolson \cite{chwolson1924} considered the particular case when the source, the lens and the observer are aligned and noticed the possibility of observing a luminous ring when a far source undergoes the lensing effect by a massive star. {In 1936, after the insistence of Rudi Mandl, Einstein published a paper on science \cite{einstein1936} describing the gravitational lensing effect of one star on another, the formation of the luminous ring, today called the Einstein ring, and giving the expression for the source amplification.} However, Einstein considered this effect exceedingly curious and useless, since in his opinion, there was no hope to actually observe it.

On this issue, however, Einstein was wrong: he underestimated technological progress and did not foresee the motivations that today induce one to widely use the gravitational lensing phenomenon.
Indeed, Zwicky promptly understood that galaxies were gravitational lenses more powerful than stars and might give rise to images with a detectable angular separation.
{ In two letters published in 1937 \cite{zwicky1937a,zwicky1937b}}, Zwicky noticed that the observation of galaxy lensing, in addition to giving a further proof of the general theory of relativity, might allow observing
sources otherwise invisible, thanks to the light gravitational amplification,
thereby obtaining a more direct and accurate estimate of the lens galaxy dynamical mass. He also found that the probability to observe lensed galaxies was much larger than that of star on star. This shows the foresight of this eclectic scientist, {since the first strong lensing event was discovered only in
1979}: the double quasar QSO 0957+561 a/b \cite{qso0957+561}, shortly followed by the observation of tens of other gravitational lenses, Einstein rings and gravitational arcs.
All of that plays today an extremely relevant role for the comprehension of the evolution of the structures and the measure of the parameters of the so-called cosmological standard~model.

Actually, there are different scales in gravitational lensing, on which we shall briefly concentrate in the next sections, after a short introduction to the basics of the theory of gravitational lensing (Section 2). Generally speaking, gravitational lens images separated by more than a few tenths of arcsecs are clearly seen as distinct images by the observer. This was the case considered by Zwicky, and the gravitational lensing in this regime is called {\em strong (or macro) lensing}
 (see Section 3), which also includes distorted galaxy images, like Einstein rings or
arcs. If instead, the distortions induced by the gravitational fields on background objects are much smaller, we have the {\em weak lensing} effect (see Section 4).
On the other side, if one considers the phenomenology of the star-on-star lensing (as~Einstein did), the resulting angular distance between the images is of the order of a few $\mu$as, generally not separable by telescopes. Gravitational lensing in this regime is called, following Paczy\`nski \cite{pacz1986}, {\em microlensing}, and the observable
is an achromatic change in the brightness of the source star over time, due to the
relative motion of the lens and the source with respect to the line of sight of
the source (see Section 5). In all of these regimes, the gravitational field can be treated in the weak field approximation. Another scale on which gravitational
lensing applies is that involving black holes. In particular, when light rays
come very close to the event horizon, they are subject to strong gravitational
field effects, and thereby, the deflection angles are large. This effect is called {\em retro-lensing}, and we shall discuss it in
Section 6. The observation of retro-lensing events is of great importance also because the general theory of relativity still stands practically untested in the strong gravitational field regime (see \cite{johannsen} for a very recent review), apart from gravitational waves \cite{gwdetection}.
A short final discussion is then offered in Section 7.

\section{Basics of Gravitational Lensing}

\begin{figure}[H]
\centering
 \includegraphics[scale=0.25]{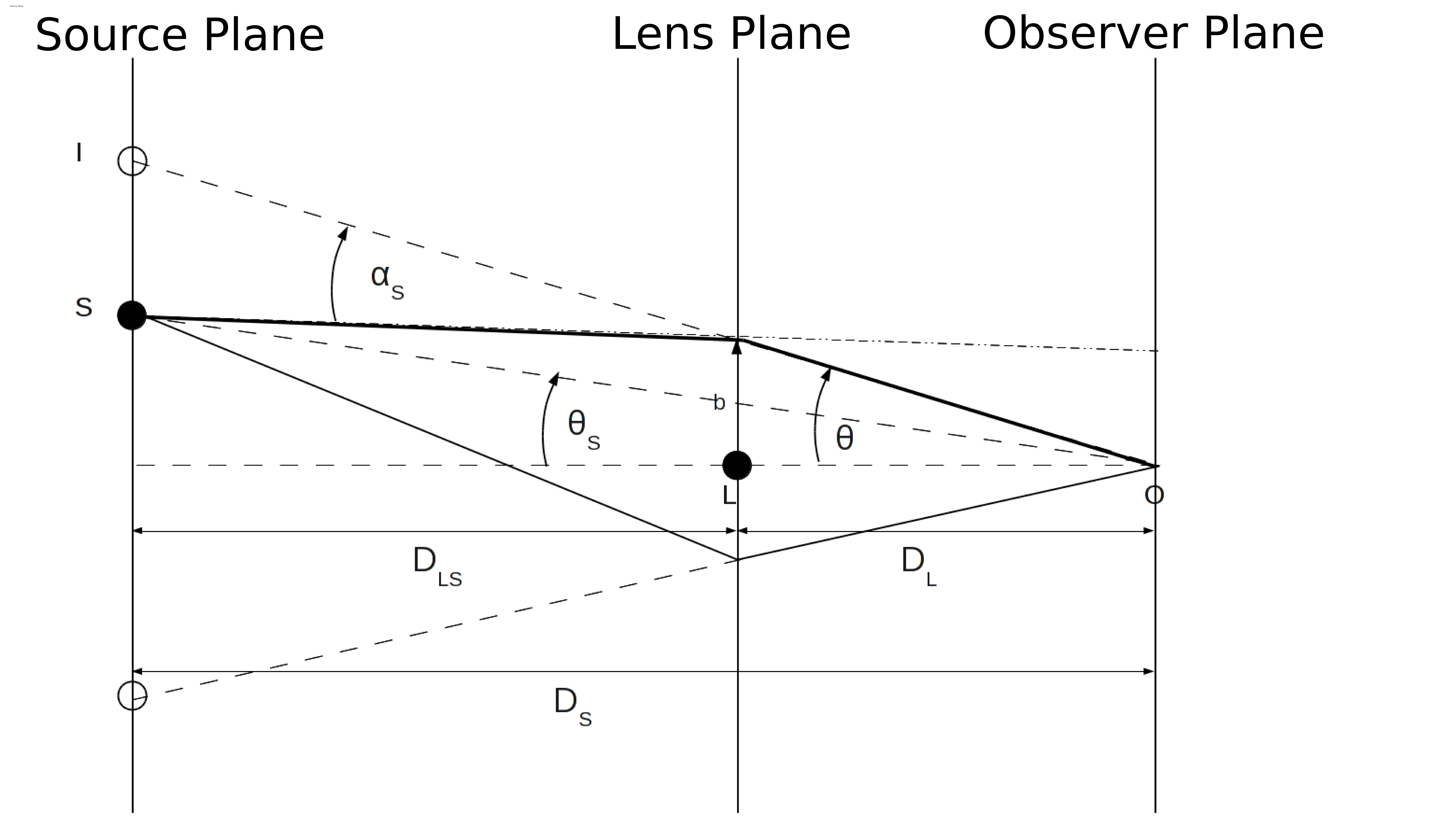}
 \caption{Schematics of the lensing phenomenon.}
 \label{FigLens}
\end{figure}

{In the general theory of relativity, light rays follow null geodesics, }
\emph {i.e.}, the minimum distance paths in a curved space-time.
Therefore, when a light ray from a far source interacts with the gravitational field due to a massive body,
it is bent by an angle approximately equal to $\alpha_S(b)=2r_s/b$. By looking at Figure \ref{FigLens}, assuming the ideal case of a thin lens and noting that $\alpha_S D_{LS}=(\theta - \theta_S) D_S$, one can easily derive the so-called lens equation:
\begin{equation}
\theta_S=\theta-\frac{\theta_E^2}{\theta},
\label{lenseq}
\end{equation}
where $\theta_S$ indicates the source position and:
\begin{equation}
\theta_E=\left(\frac{4GM}{c^2}\frac{D_{LS}}{D_S D_L}\right)^{1/2} \label{thetaE}
\end{equation}
 is the Einstein ring radius, which is the angular radius of the image when lens and source are perfectly aligned ($\theta_S=0$).
Therefore, one can see that two images appear in the source plane, whose positions can be obtained by solving Equation (\ref{lenseq}).


More generally (see for details \cite{SchneiderEhlersFalco}), the light deflection between the two-dimensional position of the source $\boldsymbol\theta_S$ and the position of the image $\boldsymbol\theta$ is given by the lens mapping equation:
\begin{equation}
\boldsymbol\theta_S=\boldsymbol\theta - \boldsymbol\nabla \phi(\boldsymbol\theta),
\label{lensing}
\end{equation}
where $\phi=2D_{LS}\Phi_N^{2D}/(D_S c^2)$ is the so-called lensing potential and $\Phi_N^{2D}$ is the two-dimensional Newtonian projected gravitational potential of the lens. We also note, in turn, that the ratio
$D_{LS}/{D_S}$ depends on the redshift of the source and the lens, as well as on the cosmological parameters $\Omega_M=\rho_M /\rho_c$ and $\Omega_{\Lambda}=\rho_{\Lambda} /\rho_c$, being
$\rho_c=3H_0^2/(8\pi G)$, $\rho_M$ and $\rho_{\Lambda}$ the critical, the matter and the dark energy densities, respectively.
{The transformation above is thus a mapping from the source plane to the image plane, and the Jacobian ${\cal J}$ of the transformation is given by:
\begin{equation}
{\cal J}=\frac{d\boldsymbol\theta_S}{d\boldsymbol\theta}={\cal A}^{-1}=
\left( \begin{array}{cc}
1-\phi_{,11} & -\phi_{,12} \\
-\phi_{,12} & 1-\phi_{,22}
\end{array} \right)
=
\left( \begin{array}{cc}
1- \kappa - \gamma_1& -\gamma_2 \\
 -\gamma_2 & 1 - \kappa + \gamma_1
\end{array} \right),
\end{equation}
where the commas are the partial derivatives with respect to the two components of $\boldsymbol\theta$.
Here, $\kappa$ is the convergence, which turns out to be equal to $\Sigma /2\Sigma_{\rm cr}$,
where:
 \begin{equation}
 \Sigma_{\rm cr}=c^2D_S/(4\pi GD_L D_{LS})
 \label{sigmacr}
 \end{equation}
 is the critical surface density,
$\boldsymbol\gamma=(\gamma_1 , \gamma_2)$ is the shear and ${\cal A}$ is the magnification matrix.
}
Thus, the previous equations define the convergence and shear as second derivatives of the potential, \emph {i.e.},
\begin{equation}
\kappa=\frac{1}{2}(\partial_1\partial_1 + \partial_2\partial_2)\phi=\frac{1}{2}\nabla^2\phi, \hspace{3mm} \gamma_1=(\partial_1\partial_1 - \partial_2\partial_2)\phi, \hspace{3mm} \gamma_2=\partial_1\partial_2 \phi .
\end{equation}
From the above discussion, it is clear that gravitational lensing may allow one to probe the total mass distribution within the lens system, which reproduces the observed image configurations and distortions. This, in turn, may allow one to constrain the cosmological parameters, although this is a second order effect.

\section{Strong Lensing}

{ Quasars are the brightest astronomical objects, visible even at a distance of billions of parsecs. After the identification of the first quasar in 1963 \cite{schmidt}, these objects remained a mystery for quite a long time, but today, we know that they are powered by mass accretion on a supermassive black hole, with a mass billions of times that of the Sun.
The first strong gravitational lens, discovered in 1979, was indeed linked to a quasar (QSO 0957+561 \cite{qso0957+561}), and although the phenomenon was expected on theoretical grounds, it left the astronomers astonished. The existence of two objects separated by about $6''$ and characterized by an identical spectrum led to the conclusion that they were the doubled image of the same quasar, clearly showing that Zwicky was perfectly right and that galaxies may act as gravitational lenses.
Afterwards, also the lens galaxy was identified, and it was established that its dynamical mass, responsible for the light deflection, was at least ten-times larger than the visible mass. This double quasar was also the first object for which the time delay (about 420 days) between the two images \cite{pelt}, due to the different paths of the photons forming the two images, has been measured. This has also allowed obtaining an independent estimate of the lens galaxy dynamical mass. Observations can also show four
images of the same quasar, as in the case of the so-called Einstein Cross, or
when the lens and the source are closely aligned, one can observe the Einstein
ring, e.g., in the case of MG1654-1346 \cite{chen}. The macroscopic effect of multiple images' formation is generally called strong lensing, which also consists of the formation of arcs, as those clearly visible in the deep sky field images by the Sloan Digital Sky Survey (SDSS; see, e.g., \cite{hennawi2008}). The sources of strong lensing
events are often quasars, galaxies, galaxy clusters and supernovae, whereas the
lenses are usually galaxies or galaxy clusters. The image separation is
generally larger than a few tenths of an arcsec, often up to a few arcsecs.

Over the years, many strong lensing events have been found in deep surveys of the sky, such~as the CLASS
 \cite{browne2003}, the Sloan ACS
 \cite{bolton2006}, the SDSS, one of the most successful surveys in the history of astronomy (see, e.g., \cite{sdss} and references therein), the SQLS (the Sloan Digital Sky Survey for Quasar Lens Search) \cite{oguri2006}, and so on.

Strong gravitational lensing is nowadays a powerful tool for investigation in astrophysics and cosmology (see, e.g. \cite{blandford1992,treu2010}).
As already mentioned in the previous section, strong lensing gives a unique opportunity to measure the dynamical mass of the lens object using, for example, the mass estimator $M(<R_E)=\pi \Sigma_{\rm cr}\theta_E^2$, which directly gives the mass within $R_E$, using Equation (\ref{sigmacr}) in this regime. The
result is that masses obtained in this way are almost always larger than the
visible mass of the lensing object, showing that galaxy and galaxy cluster
masses are dominated by dark matter. In any case, accurately constraining the mass
distribution of the lens system (e.g., a galaxy cluster) is a generally
degenerate problem, in the sense that there are several mass distributions that
can fit the observables; thus, the best way to solve it is to use multiple images (see, e.g., \cite{kneib1996}).

Another important application of strong lensing is the study of dark matter halo substructures. Indeed, sometimes flux ratio anomalies in the lensed quasar images are detected (see, e.g., \cite{mao1998,metcalf2001}), and while smooth mass models of the lensing galaxy may generally explain the observed image positions, the prediction of such models of the corresponding fluxes is frequently violated.
Especially in the radio band observations, since the quasar radio emitting region is quite large, the observed radio flux anomalies are explained as being due to the presence of substructures of about $10^6-10^8 {\rm ~M_{\odot}}$ along the line of sight. After some controversy regarding whether $\Lambda$CDM (cold dark matter plus Cosmological Constant) simulations predict enough dark matter substructures to account for the observations (for example, in \cite{xu2009}, some indication is found of an excess of massive galaxy satellites), more recent analysis, taking also into account the uncertainty in the lens system ellipticity, finds results consistent with those predicted by the standard cosmological model \cite{metcalf2012,xu2015}. However, at present, the list of multiply-imaged quasars observed in the radio and mid-IR bands is quite short, and further observational and theoretical work would be very helpful in this respect. Another indication of dark matter halo substructures comes from detailed analysis of galaxy-galaxy lensing. Although the results obtained are generally consistent with $\Lambda$CDM simulations, more data should be analyzed in order to get strong constraints \cite{vegetti2012,vegetti2014}. Strongly lensed quasars have been observed to show a certain variability of one image with respect to the others. This can be often attributed to microlensing (see Section 5) by the stars throughout the lens galaxy. This effect, and in particular its variation with respect to the wavelength, has provided an opportunity to study in detail the central engine of the source quasar, and the magnitude of the microlensing variability has allowed astrophysicists to constrain the stellar density in the lens galaxy \cite{kochanek2004,mosquera2011,pooley2012}.

Strong gravitational lensing may be used as a natural telescope that magnifies dim galaxies, making them easier to be studied in detail. For this reason, mass concentrations, like galaxies and clusters of galaxies, can be effectively used as cosmic telescopes to study faint sources that would not be possible to detect in the absence of gravitational lensing (see, e.g., \cite{soucail1987,stark2008}). At present, there is also an event of a high magnified supernova multiply imaged and also seen exploding again, being lensed by a galaxy in the cluster MACS J1149.6+2223 \cite{kelly2015,treu2016}.

The ultimate goal of strong lensing is not only to get information on the large-scale structure of the Universe, but also to constrain the cosmological parameters. For instance, analyzing the time delay among the lensed source images, it is possible to estimate also the value of the Hubble constant $H_0$. Indeed, the time delay is given by the difference of the light paths from the images and is inversely proportional to $H_0$, as first understood by Refsdal \cite{refsdal1964} (see also the review in \cite{jackson2007}). At present, one of the most accurate measurement of the Hubble constant using a gravitational lens is provided in~\cite{suyu2010}. There is also a project (COSMOGRAIL
) particularly devoted to the time delay measurements of doubly- or multiply-lensed quasars (see \cite{cosmograil2016} and the references therein). Moreover, the measure of both the frequency of occurrence and the redshift of multiple images in deep sky surveys may allow one to constrain the values of $\Omega_M$ and $\Omega_{\Lambda}$ in an independent way with respect to other methods, such as those coming from SN Ia or the CMB (Cosmic Microwave Background) power spectrum.
}

\section{Weak Lensing}

 In addition to the macroscopic deformations discussed in the previous section, in the deep field surveys of the sky, also arclets (\textit{i.e}., single distorted images with an elliptical shape) and weakly distorted images of galaxies, with an almost invisible individual elongation, have been detected. This effect is known as weak lensing and is playing an increasingly important role in cosmology.

{
The weak lensing's main feature is the shape
deformation of background galaxies, whose light crosses a mass distribution
(e.g., a galaxy or a galaxy cluster) that acts as a gravitational lens. Actually, as discussed in Section 2, gravitational lensing gives rise to two distinct effects on a source image: convergence, which is isotropic, and shear, which is anisotropic. In the weak lensing regime, the observer makes use of the shear, that is the image deformation (sometimes related to the galaxy orientation), while the convergence effect is not used, since the intrinsic luminosity and the size of the lensed objects are unknown. For a complete and in-depth review on the basics of weak gravitational lensing, with full mathematical details of all the most important concepts, we refer the reader to \cite{bartelmann2001}.

The first weak lensing event was detected in 1990 as statistical tangential alignment of galaxies behind massive clusters \cite{tyson1990}, but only in 2000, coherent galaxy distortions were measured in blind fields, showing the existence of the cosmic shear (see, e.g., \cite{bacon2000,kaiser2000}). Here, we remark that the weak lensing cannot be measured by a single galaxy, but its observation relies on the statistical analysis of the shape and alignment of a large number of galaxies in a certain direction.

Therefore, the game is to measure the galaxy ellipticities and orientations and to relate them to the surface mass density distribution of the lens system (generally a galaxy cluster placed in between). There are at least two major issues in weak lensing studies, one mainly relying on the theory, the other one on observations: the former concerns finding the best way to reconstruct the intervening mass distribution from
the shear field
$\boldsymbol\gamma=(\gamma_1 , \gamma_2)$, the latter with looking for the best way to determine the {\it true} ellipticity of a faint galaxy, which is smeared out by the instrumental point spread function (PSF). To solve these issues, several approaches have been proposed, which can be distinguished into two broad families: direct and inverse methods. On the theoretical side, the direct approaches are: the integral method, which consists of expressing the projected mass density distribution as the convolution of $\boldsymbol\gamma$ by a kernel (see, e.g., \cite{seitz1996}), and the local inversion method, which instead starts from the gradient of
$\boldsymbol\gamma$ (see, e.g., \cite{lombardi1999} and the references therein). The inverse approaches work on the lensing potential $\phi$ (see Equation \ref{lensing}), and they include the use of the
maximum likelihood \cite{schneider2000,han2015} or the maximum entropy methods \cite{marshall2002} to determine the most likely projected mass distribution that reproduces the shear field. The inverse methods are particularly useful since they make it possible to quantify the errors in the resultant lensing mass estimates, as, for instance, errors deriving from the assumption of a spherical mass model when fitting a non-spherical system \cite{clowe2004, corless2007}.

The inverse methods allow one also to derive constraints from external observations, such as X-ray data on galaxy clusters' strong lensing or CMB lensing. In particular, one can compare mass measurements from weak lensing and X-ray observations for large samples of galaxy clusters~ \cite{hoekstra2007}. In this respect, \cite{zhang2008} used a large sample of nearby clusters with good weak lensing and X-ray measurements to investigate the agreement between mass estimates based on weak lensing and X-ray data, as well as studied the potential sources of errors in both methods. Moreover, a combination of weak lensing and CMB data may provide powerful constraints on the cosmological parameters, especially on the Hubble constant $H_0$, the amplitude of fluctuations $\sigma_8$ and the matter cosmic density $\Omega_m$~\cite{contaldi2003, hollenstein2009}. We also mention, in this
respect, that one way to determine the fluid-mechanical properties of dark
energy, characterized by its sound speed and its viscosity apart from its
equation of state, is to combine Planck data with galaxy clustering and weak
lensing observations by Euclid, yielding one percent sensitivity on the dark
energy sound speed and viscosity \cite{majerotto2016} (see the end of this~section).

On the observational side, the first priority is to use a telescope with a wide field of view, appropriate to probe the large-scale structure distribution at least of a galaxy cluster. On the other hand, it is also necessary to minimize the source of noise in the determination of the ellipticity of very faint galaxies, so that the best-seeing conditions for a ground-based telescope or, better, a space-based instrument, are extremely useful.

Very promising results have been obtained with the weak lensing technique so far, as, for example, the best measure, until today, of the existence and distribution of dark matter within the famous Bullet cluster \cite{clowe2006} (actually constituted by a pair of galaxy clusters observed in the act of colliding). Astronomers found that the shocked plasma was almost entirely in the region between the two clusters, separated from the galaxies. However, weak lensing observations showed that the mass was largely concentrated around the galaxies themselves, and this enabled a clear, independent measurement of the amount of dark matter.

With the major aim to map, through the weak lensing effect, the mass distribution in the Universe and the dark energy contribution by measuring the shape and redshift of billions of very far away galaxies (for a review, see \cite{amendola}), the European Space Agency (ESA) is planning to launch the Euclid satellite in the near future. Also ground-based telescopes will allow one to detect an enormous number of weak and strong lensing events. An example is given by the LSST (Large Synoptic Survey Telescope) project, located on the Cerro Pach\'on ridge in north-central Chile, which will become operative in 2022. Its 8.4-meter telescope uses a special three-mirror design, creating an exceptionally wide field of view, and has the ability to survey the entire sky in only three nights. The effective number density $n_{\rm eff}$ of weak lensing galaxies (which is a measure of the statistical power of a weak lensing survey) that will be discovered by LSST is, conservatively, in the range of \mbox{18--24 arcmin$^{-2}$ }(see Table 4 in \cite{chang2013}).
The very large \mbox{(about $1.5\times 10^4$ square degrees)} and deep survey of the sky that will be performed by Euclid will allow astrophysicists to address fundamental questions in physics and cosmology about the nature and the properties of dark matter and dark energy, as well as in the physics of the early Universe and the initial conditions that provided the seeds for the formation of cosmic structure. Before closing this section, we also mention that strong systematics may be present in weak lensing surveys. For example, the intrinsic alignment of background sources may mimic to an extent the effects of shear and may contaminate the weak lensing signal. However, these systematics may be controlled if also the galaxy redshifts are acquired, and this fully removes the unknown intrinsic alignment errors from weak lensing detections (for~further details, see \cite{Heymans2003,Valageas2014}).
}

\section{Microlensing}

\begin{figure}[H]
\centering
\includegraphics[scale=0.26]{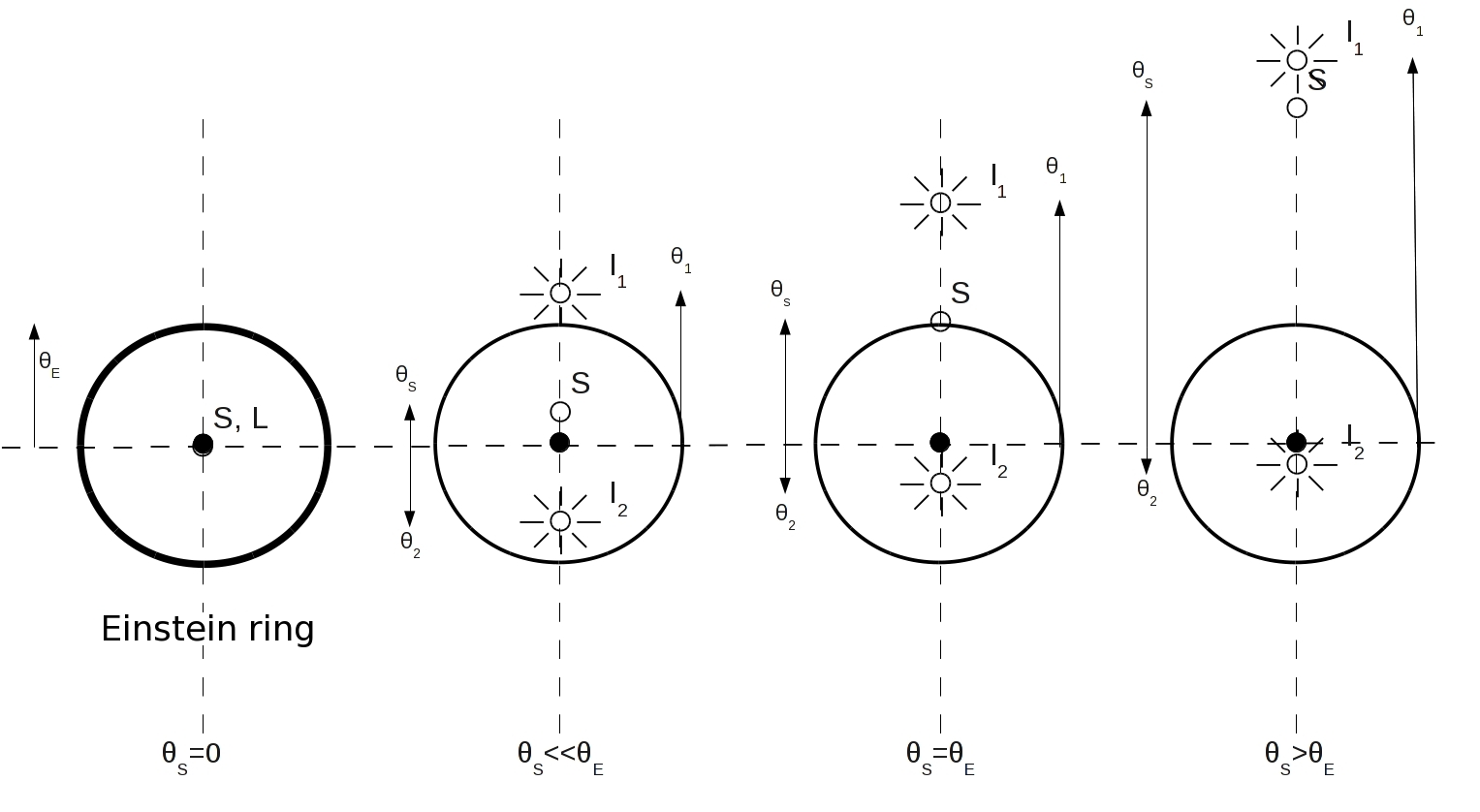}
\caption{Angular positions of the primary ($I_1$) and secondary ($I_2$) images for four different values of the source impact parameter $\theta_S$ in the case of a point-like source.}
 \label{fig_micro_lensing}
\end{figure}

Let us consider now the microlensing scale of the lensing phenomenon that occurs when $\theta_E$ is smaller than the typical telescope angular resolution, as in the case of stars lensing the light from background stars (for a review on gravitational microlensing and its astrophysical applications, we refer to, e.g., \cite{mao2012}). As is clear from the discussion in Section 2, by solving the lens Equation (\ref{lenseq}), one can determine the angular positions of the primary ($I_1$) and secondary ($I_2$) images. In~Figure~\ref{fig_micro_lensing}, these positions are shown for four different values of the impact parameter
$\theta_S$ in the case of a point-like source.
If the source and the lens are aligned (first panel on the left), the circular symmetry of the problem leads to the formation of a luminous annulus having radius $\theta_E$ around the lens position.
Otherwise, increasing the $\theta_S$ value, the secondary image gets closer to the lens position, while the primary image drifts apart from it, and in the limit of $\theta_S \gg \theta_E$, the microlensing phenomenon tends to disappear.
However, observing multiple images during a microlensing event is practically impossible with the present technology. For instance, in the case in which the phenomenon is maximized, corresponding to the perfect
alignment, for a star in the galactic bulge (about $8$ kpc away), one has $\Delta\theta=2\theta_E\simeq 1$ $\mu$as, which is well below the angular resolving power, even of the Hubble Space Telescope (about $43$ mas 
 at 500 nm); {see, e.g., http://www.coseti.org/9008-065.htm}.

\begin{figure}[H]
\centering
\includegraphics[scale=0.35]{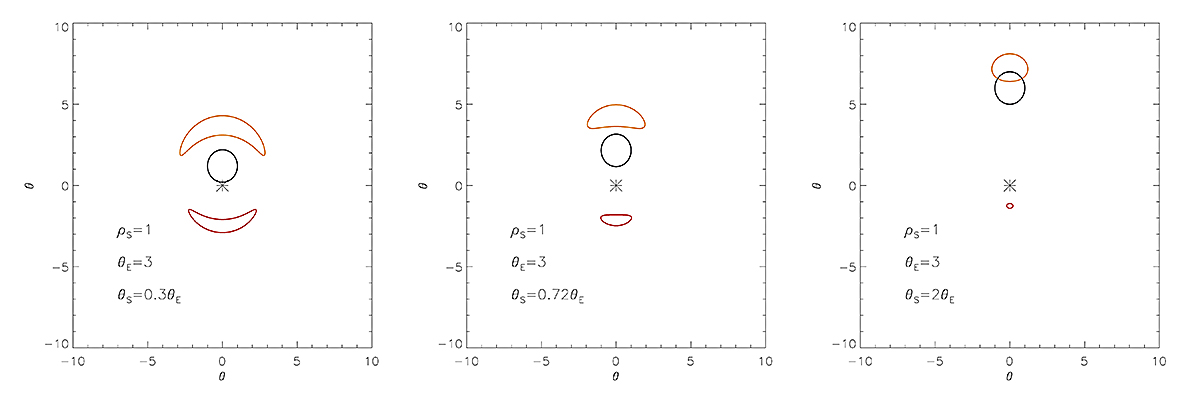}
\caption{As in Figure \ref{fig_micro_lensing}, but with a conformal transformation of the source boundary, considered extended and with radius $\rho_S$, by a point-like lens. Each point of the source disk behaves as a point-like source. The black circle represents the source disk, while the red and yellow arcs are the deformed primary and secondary images.}
 \label{source_contour}
\end{figure}

When a source is microlensed, its images do not have the same luminosity; therefore, the observer receives a total flux (or magnitude) different from that of the unlensed source.
The flux difference can be described very simply in terms of the light magnification and the law of the conservation of the specific intensity $I$,
which represents the energy, with frequency in the range $d\nu$ crossing the surface $dA$ during the time interval $dt$
in the solid angle $d\Omega$ around the direction orthogonal to the surface. Indeed, the light specific intensity turns out to be conserved in the absence of absorption phenomena,
interstellar scattering or Doppler shifts. This is also a consequence of Liouville's theorem, which claims that the density of states in the phase space
remains constant if the interacting forces are non-collisional (and gravitation fulfills this condition due to its weak coupling constant), and the propagating medium is approximately transparent (as is the case for interstellar space).
This effect can produce
a magnification or a de-magnification of the images of an extended light source (see~Figure~ \ref{source_contour}).
If the image is magnified, it means that it certainly subtends a wider angle with respect to that subtended by the source in the absence of the lens.
In microlensing, the source disk size should not be neglected in general. Within the framework of the finite source
approximation for a source with flux $F_S$ and assuming ${\theta}_E \leq {\theta}_S$, one can show that the magnification $A$ of an image at angular position $\theta$ is given by
$(1-\theta_E^4/\theta^4)^{-1}$. As a consequence, the observed flux corresponds
to $F=A F_S$. Of course, when the source star disk gradually moves away from the line of sight, the magnification decreases, and the unlensed $F_S$ flux is then recovered.

As already anticipated, the observer cannot see, in the microlensing case, well-separated images, but, instead, detects a single image made by the overlapping
of the primary and the secondary images. In this case, one can easily obtain the classical magnification factor $A$ by summing up the individual magnifications,
\emph {i.e.}, $A=(u^2+2)/\sqrt{u^2(u^2+4)}$, where $u={\theta}_S/{\theta}_E$ is the impact factor. If~there is a relative movement between the lens and the source,
$u$ changes with time, and a standard Paczy\`nski curve \cite{pacz1986} does emerge.

An important role in gravitational microlensing is played by the {\it caustics}, the geometric loci of the points belonging to the lens plane where the light magnification of a point-like source becomes infinite, and by the corresponding {\it critical} curves in the source plane.
In the case of a single lens, the caustic is a point coinciding with the lens position; therefore, the magnification diverges when the impact
parameter approaches zero. However, real sources are not point-like, so we always have finite magnifications that can be calculated by an average procedure:
\begin{equation}
\langle A \rangle = \frac{\int A({\bf y})I({\bf y})d^2 y}{\int I({\bf y})d^2 y} ,
\end{equation}
where $A({\bf y})$ is the point-like source magnification, $I({\bf y})$ is the brightness profile of the stellar disk (the limb darkening profile) and the integral is extended over the source star disk.

\begin{figure}[H]
\includegraphics[scale=0.6]{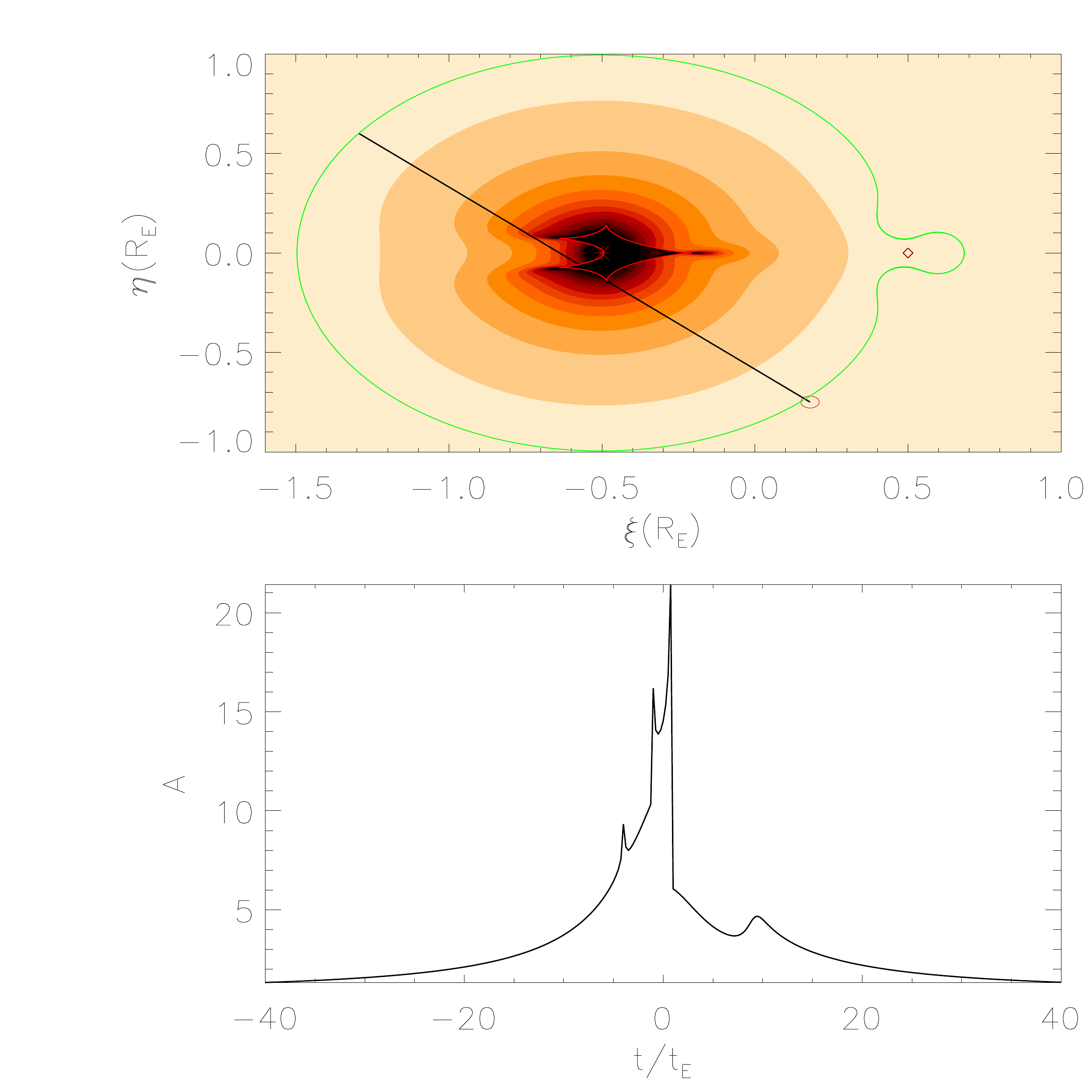}
\caption{Magnification map for a binary lens system characterized by two objects separated by a projected distance of $1R_E$ and mass ratio $q=0.01$.
The green and red closed lines indicate the critical and caustic curves obtained by solving the lens equation in Equation (1).
 The black line indicates the trajectory of the source star, which has a radius of $0.03R_E$. The simulated light curve is shown in the lower panel.}
 \label{mappa}
\end{figure}

Observations show that about half of all stars are in binary systems, and moreover, thousands of exoplanets are being discovered around their host stars by different techniques and instruments. Therefore, it is worth considering binary and multiple systems as lenses in microlensing observations. In this case, the lens equation, obviously, becomes more complicated, but it can still be solved by numerical methods in order to obtain the magnification map where caustics take on distinctive shapes depending on the specific geometry of the system. In Figure \ref{mappa}, we show the magnification map and the resulting light curve for a simulated microlensing event due to a binary lens with mass ratio $q=M_1/M_2\simeq0.01$ (e.g., a solar mass as the primary component and a Jupiter-like planet as the secondary one).
In these cases, the resulting light curve may be rather different with respect to the typical Paczy\`nski one, depending on the system parameters.
The study of these anomalies in the microlensing light curves behavior is becoming more and more important nowadays,
since it allows one to estimate some of the parameters of the lensing system (see, e.g., \cite{Perryman14}). The main advantage of this technique, compared to the other methods adopted by the exoplanets hunters (e.g., radial velocity, direct imaging, transits), is the possibility to detect even very small planets orbiting their own star at enormous distances from Earth. {{It also} allows one to discover the so-called free-floating planets (FFPs), otherwise hardly detectable \cite{sumi}.}
By studying the PA99-N2 microlensing event, detected in 1999 by the French-British collaboration POINT-AGAPE
 \cite{an}, Ingrosso \emph {et al}. \cite{Ingrosso2009} revealed in 2009 that the anomaly observed was compatible with the
presence of a super-Jupiter with a mass of $\simeq 5 M_J$ around a star lying in the
Andromeda galaxy (see Figure 5), thus finding the first putative
exoplanet in another galaxy.

\begin{figure}[H]
 \centering
 \includegraphics[scale=0.2]{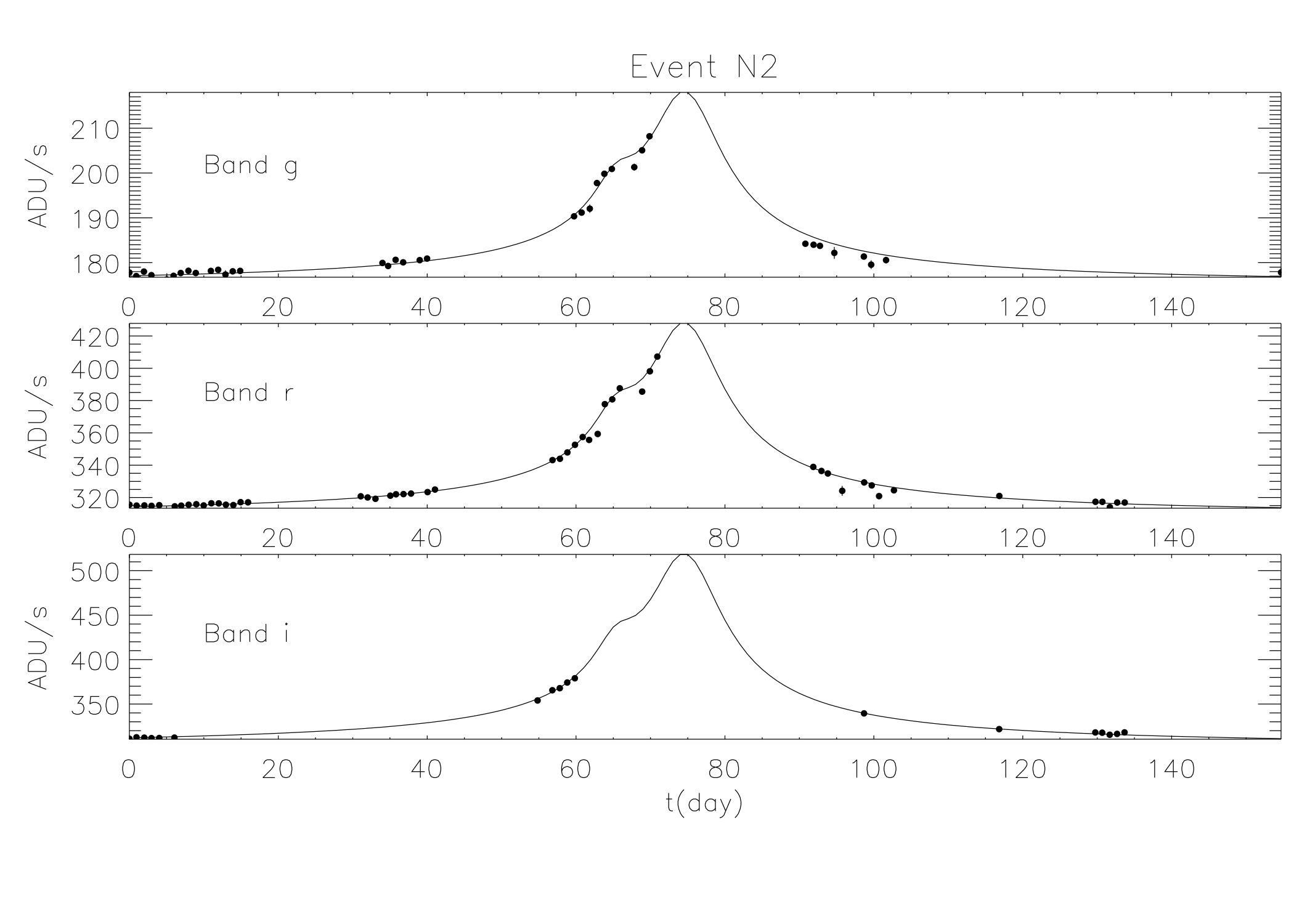}
\vspace{-12pt}
 \caption{Light curve in three different bands (g, r and i) of the PA99-N2 event detected in 1999 toward the Andromeda galaxy. }
 \label{PA99N2}
\end{figure}


\subsection{Astrometric Microlensing}

During an ongoing microlensing event, the centroid of the multiple images and the source star positions move in the lens plane giving rise to
a phenomenon known as astrometric microlensing (see, e.g., \cite{dominik2000,lee2010} and the references therein).
In the simplest case of a point-like lens, for a source at angular distance $\theta_S$, the position $\theta$ of the images with respect
to the lens can be obtained by solving the lens Equation (\ref{lenseq}).
Since the Einstein radius $R_E=D_L\theta_E$ defines the scale length on the lens plane, the lens equation reads:
\begin{equation}
d^2-d d_S -R_E^2=0,
\end{equation}
where $d_S$ and $d$ are the linear distances, in the lens plane, of the source and images from the gravitational lens, respectively.
Moreover, using the dimensionless source-lens distances $u=\theta_S/\theta_E$ and $\tilde{u}=\theta/\theta_E$, the previous relation can be further simplified as:
\begin{equation}
\tilde{u}^2-u\tilde{u} -1=0.
\end{equation}

Denoting with $u_+$ and $u_-$ the solutions of this equation, one notes that, in the lens plane, the $+$ image resides
always outside the circular ring centered on the lens position with radius equal to the Einstein angle,
while the $-$ image is always within the ring. As the source-lens distance increases, the $+$ image approaches the source position, while the $-$ one (becoming fainter) moves towards the lens location. For a source moving in the lens plane with transverse velocity $v_{\perp}$ directed along the $\xi$ axis ($\eta$ is perpendicular to it),
the projected coordinates of the source result in being:
\begin{equation}
\xi(t)=\displaystyle{\frac{t-t_0}{t_E}},~~~\eta(t)=\displaystyle{u_0}.
\label{eqcoordinatescenter}
\end{equation}
where $t_E=R_E/v_{\perp}$ and $u_0$ is the impact parameter
 (in this case, lying on the $\eta$ axis).
Since $u^2=\xi^2+\eta^2$ is time dependent, the two images move in the lens plane during the gravitational lensing event.

By weighting the $+$ and $-$ image position with the associated magnification \cite{walker1995}, one gets:
\begin{equation}
\bar{u}\equiv\frac{\tilde{u}_+\mu_+ +\tilde{u}_-\mu_-}{\mu_+ +\mu_-}=\frac{u(u^2+3)}{u^2+2}.
\label{centroidpair}
\end{equation}

Finally, the observable is defined as the displacement of the centroid with respect to the source,
\begin{equation}
\Delta\equiv\bar{u}-u=\frac{u}{2+u^2}.
\label{shiftmodulus}
\end{equation}

Note that the centroid shift may be viewed as a vector:
\begin{equation}
{\bf \Delta}=\frac{{\bf u}}{2+u^2}
\label{shiftarray}
\end{equation}
with components along the axes:
\begin{equation}
\Delta_{\xi}=\displaystyle{\frac{\xi(t)}{2+u^2}},~~~ \Delta_{\eta}=\displaystyle{\frac{u_0}{2+u^2}}.
\label{shiftcomponents}
\end{equation}

Here, we remind that all of the angular quantities
are given in units of the Einstein angle $\theta_E$, which, for a source at distance $D_S \gg D_L$, results in being:
\begin{equation}
\theta_E\simeq2\left(\frac{M}{0.5 {\rm ~M_{\odot}} }\right)^{1/2}\left(\frac{D_L}{{\rm kpc} }\right)^{-1/2}~{\rm mas},
\end{equation}
which fixes the scale of the phenomenon.

It is straightforward to show (see \cite{walker1995}) that during a microlensing event,
the centroid shift $\Delta$ traces (in the
$\Delta_{\xi}, \Delta_{\eta}$ plane) an ellipse centered in the point
$(0,b)$. The ellipse semi-major axis $a$ (along $\Delta_{\eta}$) and
semi-minor axis $b$ (along $\Delta_{\xi}$) are:
\begin{equation}
a=\frac{1}{2}\frac{1}{\sqrt{u_0^2+2}},~~~b=\frac{1}{2}\frac{u_0}{u_0^2+2}.
\label{axes}
\end{equation}
Then, for $u_0 \rightarrow \infty$, the ellipse becomes a circle with radius $1/(2u_0)$, while it
degenerates into a straight line of length $1/\sqrt{2}$ for $u_0$ approaching zero.
Note also that Equation (\ref{axes}) implies:
\begin{equation}
u_0^2=2 (b/a)^2 \left[1-(b/a)^2\right]^{-1} ,
\label{u0fromaxes}
\end{equation}
so that by measuring $a$ and $b$, one can determine the event impact parameter $u_0$.

As observed in \cite{dominik2000}, ${\bf \Delta}$
falls more slowly than the magnification, implying that the centroid shift may be an interesting observable also
for large source-lens distances, \emph {i.e.},
far from the light curve peak. In fact,
in astrometric microlensing, the threshold impact parameter $u_{\rm th}$ (\emph {i.e.}, the value of
the impact parameter that gives an astrometric centroid signal larger than a certain
quantity $\delta_{\rm th}$)
is given by \mbox{$u_{\rm th}=\sqrt{T_{\rm obs} v_{\perp}/(\delta_{\rm th} D_L)}$, where $T_{\rm obs}$}
is the observing
time and $v_{\perp}$ the relative velocity of the source
with respect to the lens. For example, the
Gaia satellite should reach an astrometric precision $\sigma_G\simeq 300$ $\mu$as
(for objects with visual magnitude $\simeq 20$) in five years of observation
\cite{eyer}. Then, assuming a threshold centroid shift $\delta_{\rm th}\simeq \sigma_G$,
one has $u_{\rm th}\simeq 60$ for a lens at a distance of $0.1$
kpc and transverse velocity $v_{\perp} \simeq 100$ km s$^{-1}$.
For comparison, the threshold impact parameter for a ground-based photometric observation
is $\simeq 1$. Consequently, the cross-section for an astrometric microlensing
measurement is much larger than the photometric one, since it scales as $u_{\rm th}^2$.
Hence, in the absence of finite-source and blending effects,
by measuring $a$ and $b$, one can directly estimate
the impact parameter $u_0$.

A further advantage of the astrometric microlensing is that some events can be predicted in advance
\cite{paczinsky1996}. In fact, by studying in detail the characteristics of stars with large proper
motions, Proft~\emph {et al}. \cite{proft2011} identified tens of candidates to measure astrometric
microlensing by the Gaia satellite, an European Space Agency (ESA) mission
that will perform photometry, spectroscopy and high precision astrometry (see \cite{eyer}).

\subsection{Polarization and Orbital Motion Effects in Microlensing Events}

Gravitational microlensing observations may also offer a unique tool to study the atmospheres
of far away stars by detecting a characteristic polarization signal
\cite{pol1}.
In fact, it is well known that the light received from stars is linearly polarized by the photon
scattering occurring in the stellar atmospheres. The mechanism is particularly effective for the hot stars
(of A or B type) that have a free electron atmosphere giving rise to a polarization degree increasing
from the center to the stellar limb \citep{chandra}.
By a minor extent, polarization may be also induced in main sequence F or G stars by the scattering of
star light off atoms/molecules and in evolved, cool giant stars by photon scattering on dust grains
contained in their extended envelopes.

{Following the approach in \cite{chandra}, the polarization $P$ in the
direction making an angle $\chi=\arccos(\mu)$ with the normal to the star surface is
$P(\mu) = [I_r(\mu)-I_l(\mu)]/[I_r(\mu)+I_l(\mu)]$, where $I_l(\mu)$ is the intensity
in the plane containing the line of sight and the normal, and $I_r(\mu)$ is
the intensity in the direction perpendicular to this plane.
Here, $\mu = \sqrt{1-(r/R)^2}$, where $r$ is the distance of a star disk element from the center
and $R$ the star radius, and we are assuming that light propagates in the direction ${\bf r \times l}$.

For isolated stars, a polarization signal has been measured only for the Sun
for which, due to the distance, the projected disk is spatially resolved.
Instead, when a star is significantly far away and can be considered as point-like, }
only the polarization $\langle P \rangle$ averaged over the stellar disk can be measured, and
usually $\langle P \rangle =0$, since the flux from each stellar disk element is the same.
A net polarization of the light appears if a suitable asymmetry in the stellar
disk is present (caused by, e.g., eclipses, tidal distortions, stellar spots, fast rotation, magnetic fields).
In the microlensing context, the polarization arises since
different regions of the source star disk are magnified differently during the event. Indeed, during an
ongoing microlensing event, the gravitational lens scans the disk of the background star,
giving rise not only to a time-dependent light magnification, but also to a time-dependent
polarization.

{This effect (see also \cite{simmons2002}) is particularly relevant in the microlensing events where:
(1) the magnification turns out to be significant;
(2) the source star radius and the lens impact parameter are comparable;
(3) the source star is a red giant, characterized by a rather low surface temperature
($T\leq 3000\,K$), around which the formation of dust grains is possible. This occurs beyond the distance $R_h$
from the star center at which the gas temperature in the stellar wind
becomes lower than the grain sublimation temperature
($\simeq 1400$ K).
The intensity of the expected polarization signal relies on the dust grain optical depth $\tau$ and
can reach values of 0.1\%--1\%, which could be reasonably observed using, for example, the
ESO VLT
 telescope (see \cite{ingrossopol2015}).
In Figure \ref{curvepol}, we show some typical polarization curves, expected in {\it bypass} (continuous curves)
and {\it transit} events (dashed curves), in which the lens trajectory approaches or passes through
the source regions where the dust grains are present. In~Figure \ref{pol2},
the distribution of the peak polarization values (given in percent) as a function of the intrinsic source star color index $(V-I)_{\rm int}$ (\emph {i.e.}, the de-reddened color of the unlensed source star)
is shown for a sample of OGLE
-type microlensing events generated by a synthetic stellar catalog simulating the bulge stellar population.
As one can see, red giants with $(V-I)_{\rm int}\leq 3$, which corresponds to the events inside the regions delimited by dashed lines, have $P_{\rm max}\leq 1$ percent values. These are the typical events observed by the OGLE-III microlensing campaign.
There are, however, a few events with $1 \leq P_{\rm max} \leq 10$ percent, characterized by $(V-I)_{\rm int}\geq 3$, corresponding to source stars in the AGB phase. These stars, which are rather rare in the galactic bulge, have not been sources of microlensing events observed in the OGLE-III campaign, but they are expected to exist in the galactic bulge.
In this respect, the significant increase in the event rate by the forthcoming generation of microlensing surveys towards the galactic bulge, both ground-based, like KMTNet
~ \cite{Henderson2014}, and space-based, like EUCLID
 \cite{Penny2013} and WFIRST
 \cite{Yee2014}, opens the possibility to develop an alert system able to trigger polarization measurements in ongoing microlensing events.

Another way to study
the atmosphere of the source star is to analyze the amplification curve and look
for dips and peaks, typically due to the presence of stellar spots
on the photosphere of the star~\cite{2000ApJ...529...69H,2002MNRAS.335..539H}. These features may be easily confused, however, with the signatures of a binary lensing system.
When the source star has a relevant rotation motion during the
lensing event, there is the possibility to really detect the stellar spots on
the source's surface and to estimate the rotation period of the
star~\cite{2015MNRAS.453.2017G}. A new generation of networks of telescopes
dedicated to microlensing surveys, like KMTNet~\cite{2010SPIE.7733E..3FK}, will
provide high-precision and high-cadence photometry that will enable us to observe
spots on the source's surface. We remark that also multicolor observations of
the event would help to disentangle the aforementioned degeneracy, as the ratio between
the brightness of the spot and the surrounding photosphere strongly depends on
the frequency of the observation. It has been shown that stellar spots can be
detected also through polarimetric observations of microlensing caustic-crossing
events~\cite{2015MNRAS.452.2587S}.

Under certain circumstances, binary lens systems are characterized by the
\emph{close-wide degeneracy}: if the two objects are separated by a projected
distance \(s\) or \(1/s\), the resulting caustics have the same structure, and
also, the observed light curves will appear the
same~\cite{1999A&A...349..108D,2005MNRAS.356.1409A}. This happens, for example,
in systems with small mass ratio \(q\), like planetary
systems~\cite{1998ApJ...500...37G}. It is possible to resolve this degeneracy
in the case of short-period binary lenses, the so-called \emph{rapidly rotating
 lenses}, as the orbital motion induces repeating features in the amplification
curve that can be exploited to estimate important physical parameters of the
lensing systems, including the orbital period, the projected separation and the
mass~\cite{2011MNRAS.417.2216P,2014MNRAS.438.2466N}.

\begin{figure}[H]
 \centering
 \includegraphics[scale=0.17]{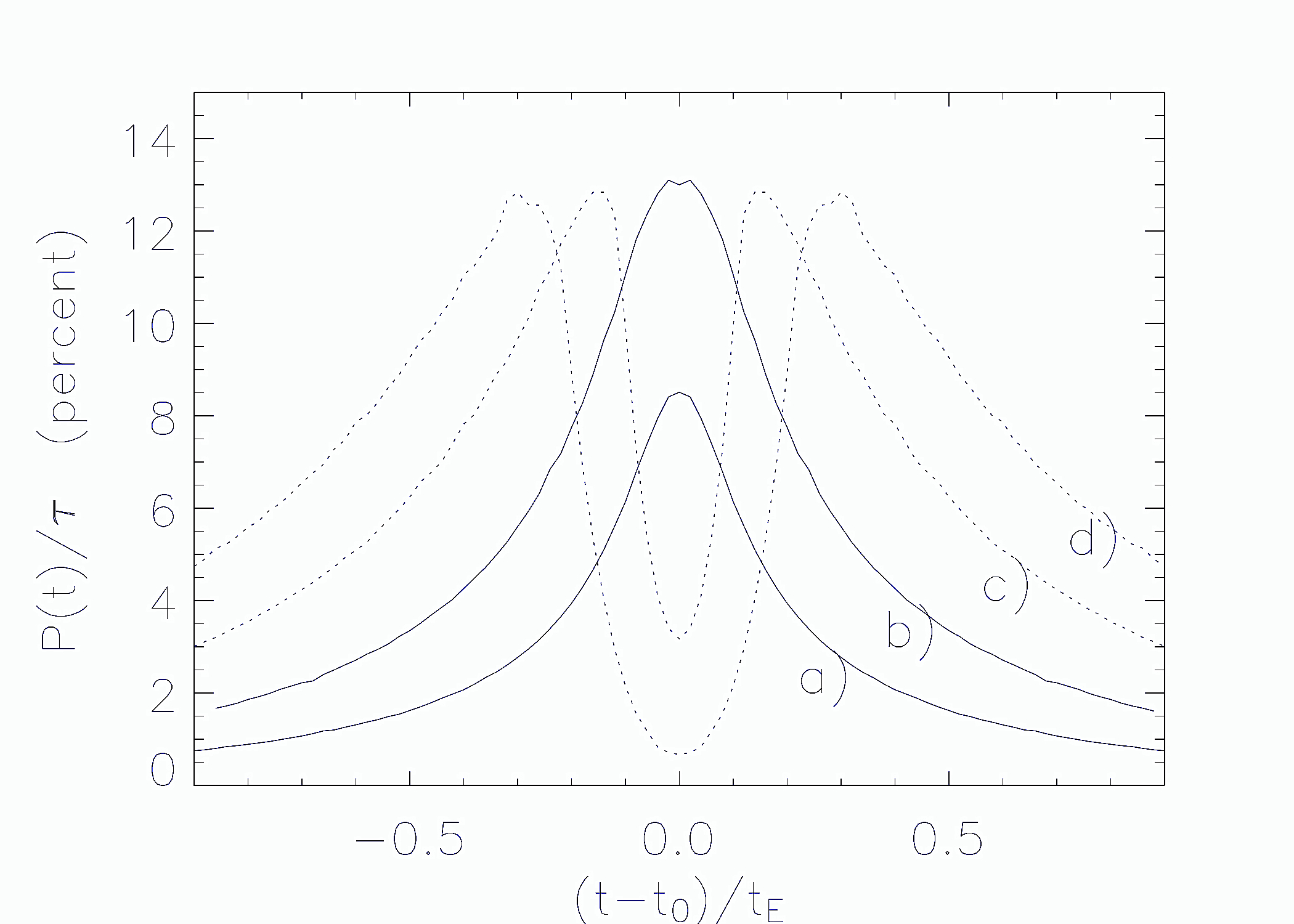}
 \caption{Assuming parameter values $u_0=0.09$, $t_E=60$ day and
$R_h/R_S = 5$, the $P(t)/\tau$ polarization curves are shown as a function of
$(t-t_0)/t_E$, for increasing values of $R_h/u_0 = 0.35, \ 0.75, \ 1.5, \ 2.5$, corresponding to the
curves labeled (a), (b) (c) and (d), respectively. Continuous curves, (a) and (b), are {\it bypass} events;
dotted lines, (c) and (d), are {\it transit} events. Here, $R_h$ is the minimum distance for the formation of dust grains, and $R_S$ is the source star radius.}
 \label{curvepol}
\end{figure}

\begin{figure}[H]
 \centering
 \includegraphics[scale=0.7]{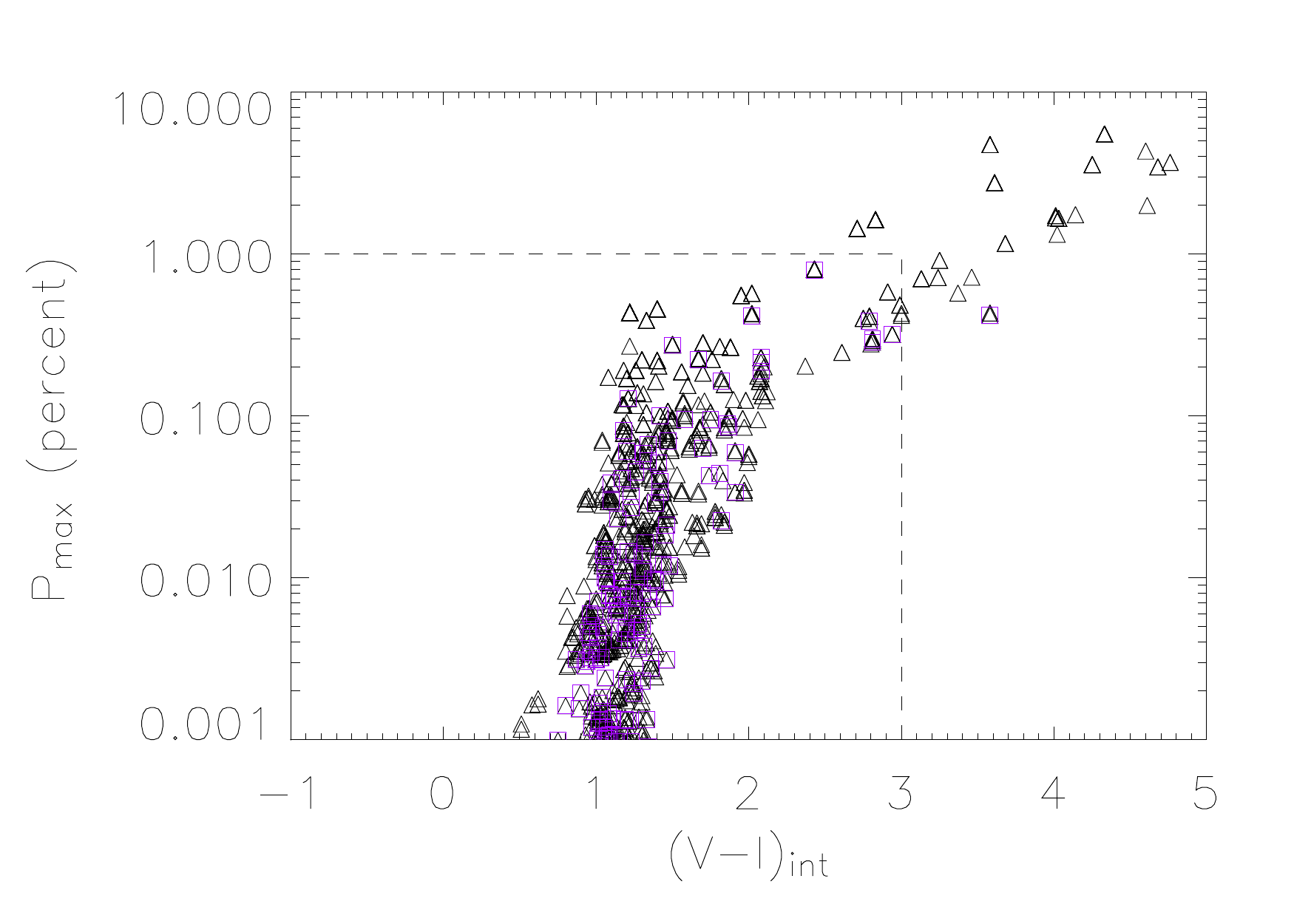}
 \caption{Distribution of the peak value of the polarization signal as a function of the intrinsic color index $(V-I)_{\rm int}$ of the source star for simulated {\it transit} (triangles) and {\it bypass} (purple squares) events.
The dashed
lines indicate the region in which the events observed by the OGLE
 Collaboration
are expected to lie.}
 \label{pol2}
\end{figure}

\section{Retro-Lensing: Measuring the Black Hole Features}

Gravitational lensing at the scales considered in the previous sections can be treated in the weak gravitational field approximation of the general theory of relativity, since in those cases, photons are deflected by very small angles. This is not the case when one considers black holes, for which it may happen that photons get very close to the event horizon of these compact objects.

Black holes are relatively simple objects. The
\emph{no-hair theorem} postulates that they are completely described by only three
parameters: mass, angular momentum (generally indicated by the spin parameter $a$) and electric charge; any other information
(for which {\it hair} is a metaphor) disappears behind the event horizon, and it is therefore inaccessible to external observers.
Depending on the
values of these parameters, black holes can be classified into Schwarzschild black holes (non-rotating and non-charged),
Kerr black holes (rotating and non-charged), Reissner--Nordstr\"om black holes (non-rotating and charged) and Kerr--Newman black holes (rotating and charged).

{Even though they appear so simple, black holes are mathematically complicated to describe (see, e.g., \cite{chandraBH}). Nowadays, we know that black holes are placed at the center of the
majority of galaxies, active or not, and in many binary systems emitting X-rays.
Moreover, they are the engine of gamma-ray bursts (GRBs) and play an essential
role in better understanding stellar evolution, galaxy formation and evolution, jets and, in the end, the nature of space and time. One goal
astrophysicists have been pursuing for a long time is to probe the immediate
vicinity of a black hole with an angular resolution as close as possible to the size
of the event horizon. This kind of observations would give a new opportunity to
study strong gravitational fields, and as we will see at the end of this
section, we think we are very close to reaching this goal.
}

{
How do we measure the mass, angular momentum and electric charge of a black hole? One
possibility, rich with interesting consequences, was suggested by Holz and Wheeler
\cite{holzwheeler}, who considered a phenomenon that was already known to be possible around black
holes. They used the Sun as the source of light rays and a black hole far from the
solar system. As shown in Figure~\ref{retro-lensing}, some photons would have the
right impact parameter to turn around the black hole and come back to
Earth. Other photons, with a slightly smaller impact parameter, can even
rotate twice around the black hole, and so on.
A series of concentric rings should then appear if the observer, the Sun and the black hole are perfectly aligned.
The two authors also suggested to do a survey and look for concentric
rings in the sky in order to discover black holes. Unfortunately, there are two problems with this
idea. First, it is unlikely that the Sun, Earth and a black hole are perfectly aligned,
and in any case, Earth moves around the Sun, so that the alignment can occur only for a short time interval.
 The second and most important problem is that the
retro-image of the Sun is so dim, that even using the Hubble Space Telescope (HST), only
a black hole with a mass larger than \(10~M_{\odot}\) within \(0.01\)~pc from
the Earth could be observed with the proposed technique. Moreover, we already know that such an object cannot be so close to the solar system without causing observable perturbations in the planet~orbits.
}

\begin{figure}[H]
 \includegraphics[scale=0.4]{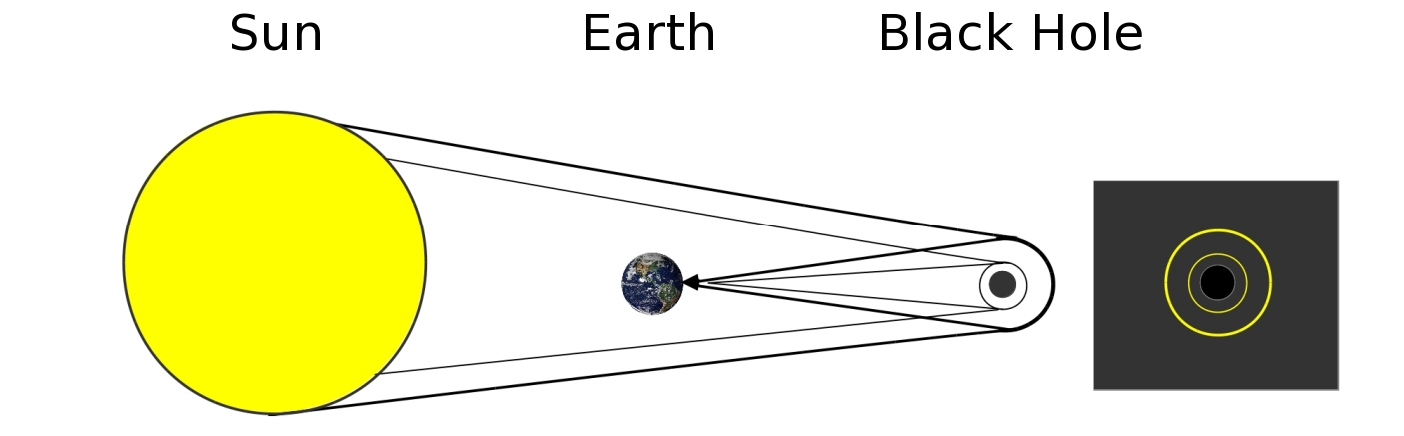}
 \caption{Retro-lensing of the Sun light by a black hole as seen from Earth. On the
  right-hand side, the series of rings around the black hole's event horizon
  an observer would observe in the case of perfect alignment. For clarity, only two rings are
  shown.}
 \label{retro-lensing}
\end{figure}

A better approach to test the idea proposed by Holz and Wheeler is to consider a well-known supermassive black hole and a bright star around it.
Of course, the brighter the source star, the brighter will be the retro-image.
Some of us~\cite{depaolisS2} soon proposed to consider retro-lensing
around the black hole at the galactic center, and in particular, the retro-lensing
image of the closest star orbiting around it. Indeed, it is known that at the
center of our galaxy, there is a supermassive black hole, with mass about $(4.2 \pm 0.2)\times
10^6~M_{\odot}$, identified by studying the orbits of several
bright stars orbiting around it (see \cite{gillessen,doku2014} and the references therein).
A method to determine the mass and the angular momentum of this black hole could then
be to measure the periastron or apoastron shifts of some of the stars orbiting around it. Another method to estimate the black hole spin $a$ is based on the analysis of the quasi-periodical oscillations towards Sgr A$^*$. Recently, the analysis of the data in the X-ray and IR bands have allowed some astrophysicists to find that $a=0.65\pm 0.05$ \cite{doku2014,doku2015}. However, there is a drawback in this
approach: periastron and apoastron shift of orbits depend not only on the black hole parameters, but also on how stars are distributed around the black
hole and on the mass density profile of the dark matter possibly present in the region surrounding the black hole. It is possible to
understand the difficulty of the measure by noting that the difference of
the periastron shift of the S2 star (the closest one to the black hole at the center of
our galaxy) induced by a Schwarzschild black hole or a
Kerr black hole with spin parameter $a = 1$ (and the same mass of the Schwarzschild one) is only of \(\simeq 10~\mu\)as (for the dependance of the periastron shift on the black hole spin orientation
see~\cite{iorio2012}).
Then, even if one had succeeded in
measuring the periastron shift of the closest star to the central black hole, it
would be unlikely to derive the amount of the black hole angular momentum. Our goal
could be achieved anyway by measuring the periastron shift of many stars orbiting
around the center of the Galaxy. The measure of the periastron shift could give, in turn, also an estimate of the parameters of the dark matter concentration expected to lie towards the Sgr $A^*$ region
 \cite{zak2007}, as well as to test different modifications of the general theory of relativity~\cite{zak2012,falcke2013,zak2014} (see also \cite{zak} for the constraints on $R^n$ theories by Solar System data)).} However, this is anything but easy \cite{pasp2007}. An important step forward in this direction has been provided recently by near-infrared astrometric observations of many stars around Sgr $A^*$ with a precision of about 170 $\mu$as in position and $\simeq 0.07$ mas$\cdot$ yr$^{-1}$ in velocity \cite{plewa2015}. A further improvement, hopefully in the near future, would make possible the direct detection of relativistic effects in the orbits of stars orbiting the central black~hole.

\begin{figure}[H]
 \centering
 \includegraphics[scale=0.37]{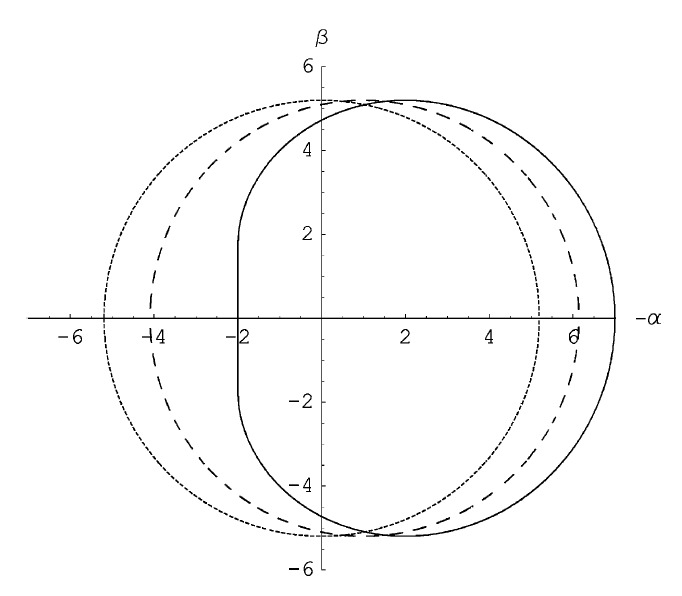}
 \caption{Retro-lensing images of a source by a
  Schwarzschild black hole (dotted circle), a Kerr black hole with spin
  parameter \(a=0.5\) (dashed line) and a maximally spinning black hole
  with \(a=1\) (continuous line). The line of sight of the observer is
  perpendicular to the spin axis of the black hole.}
 \label{retroshape}
\end{figure}

{ Retro-lensing images of bright stars retro-lensed by the black hole at the
galactic center might give an alternative method to estimate the Sgr A$^*$ black hole parameters.
Even though in general it is
difficult to calculate the retro-lensing images, since this requires
integrating with high precision the trajectories followed by the light, it is possible to numerically do these calculations not only for a
Schwarzschild black hole, but also for Kerr and Reissner--Nordstr\"om black
holes. As discussed in several papers (see, e.g., \cite{depaolis2011} and the references therein), one finds
that the shape of the retro-lensing image depends on the black hole spin
(see Figure~\ref{retroshape}), and then, in
principle, a single precise enough observation of the retro-lensing image
of a star could allow one to unambiguously estimate the parameters of the black hole in
Sgr A
$^*$. It is possible to show that also the electric charge of a Reissner--Nordstr\"om black hole can be obtained~\cite{zak2005}. In fact, although the formation of a Reissner--Nordstr\"om black hole may be problematic, charged black holes are objects of intensive investigations, and the black hole charge can be estimated by using the size of the retro-lensing images that can be revealed by future astrometrical missions. The shape of the retro-lensing (or shadow) image depends in fact also on the electric charge of the black hole,
 and it becomes smaller as the electric charge increases. The mirage size difference between the extreme charged black hole and the Schwarzschild black hole case is about 30\%, and in the case of the black hole in
Sgr A$^*$, the shadow typical angular sizes are about 52 $\mu$as for the Schwarzschild case and about 40 $\mu$as for a maximally charged Reissner--Nordstr\"om black hole. Therefore, a charged black hole could be, in principle, distinguished by a Schwarzschild black hole with RADIOASTRON
, at least if its charge is close to the maximal value. We also mention that the black hole spin gives rise also to
 chromatic effects (while for non-rotating lenses, the gravitational lensing effect is always achromatic), making one side of the image bluer than the other side \cite{depaolis2011}.
}

\begin{figure}[H]
 \centering
 \includegraphics[scale=0.2]{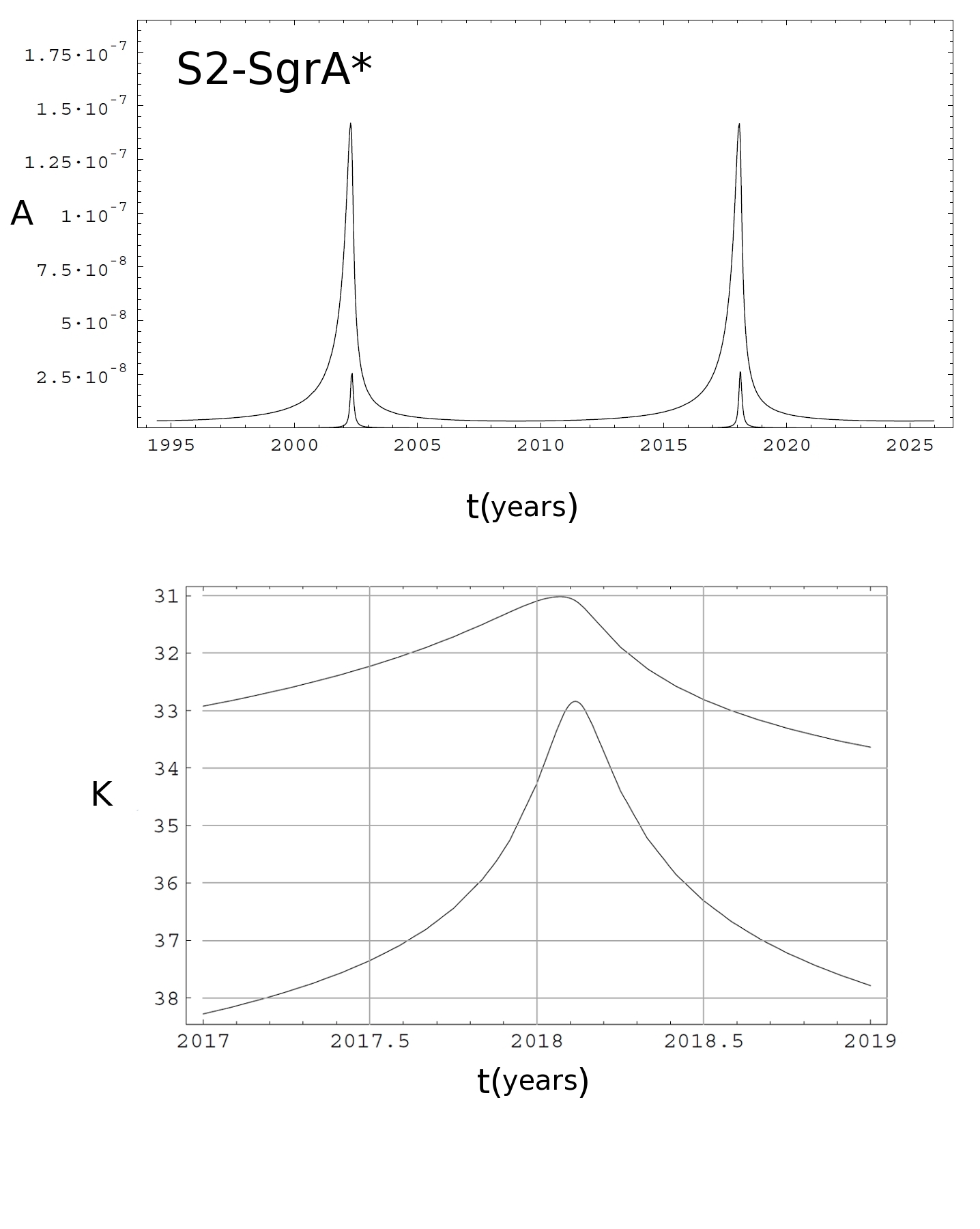}
\vspace{-43pt}
 \caption{Upper panel: amplification as a function of time for the primary (upper curve)
 and secondary (lower curve) retro-lensing images of the S2 star by the black
 hole in the Galaxy center. Lower panel: light curve in K-band magnitude of the two retro-lensing images (adapted from \cite{bozzamancini2004}).
 The standard interstellar absorption coefficient towards the Galaxy center has been assumed. }
 \label{s2sgra}
\end{figure}

Can we really hope to observe these retro-lensing images towards Sgr A$^*$?
Despite what one could think, we are not so far from this goal. The successor
of the Hubble Space Telescope, the James Webb Space Telescope (JWST), scheduled for
launch in October 2018, has the sensitivity to observe the retro-lensing
image of the S2 star produced by the black hole at the galactic center with an exposure
time of about thirty hours. In Figure~\ref{s2sgra}, we show the magnification (upper panel) and
the magnitude (bottom panel) light curves (in K band) of the retro-lensing image of the S2 star produced by the black hole at the galactic
center (see also \cite{bozzamancini2004}).
Unfortunately, JWST has not the angular resolution
necessary to provide information about the shape of the retro-lensing
image. The right angular resolution could be gained with the next generation of radio
interferometers. In fact, the diameter of the retro-lensing image around
the central black hole should be of about $30$ $\mu$as, and already in 2008, Doeleman
and his collaborators \cite{doeleman} managed to achieve an angular resolution of
about $37$ $\mu$as, very close to the required one, by using
interferometrically different radio telescopes with a baseline of
about 4500 km. Progress in this
field is so fast, that it is not hard to think we can eventually reach this aim
in the near future by, e.g., the EHT (Event Horizon Telescope) project, or by the planned
Russian space observatory, Millimetron (the spectrum-M project), or by combined observations with different interferometers, such as the Very Large Array
(VLA) and ALMA (Atacama Large Millimeter Array).

\section{Conclusions}

In the paper, we have discussed the various scales in which gravitational lensing manifests itself and that may lead us to obtain valuable information about a great variety of astronomical issues ranging from the star distribution in the Milky Way, the study of stellar atmospheres, the discovery of exoplanets in the Milky Way and also in nearby galaxies, the study of far away galaxies, galaxy clusters and black holes. Gravitational lensing, in particular in the strong and weak lensing regime, may also allow scientists to answer, in the near future, fundamental questions in cosmology related to the nature of dark matter, why the Universe is accelerating and what is the nature of the source responsible for the acceleration, which physicists refer to as dark energy.


\acknowledgments{\textbf{Acknowledgments:} The authors acknowledge the support of the TAsP (Theoretical Astroparticle Physics)
 Project funded by INFN.  Mos\`e Giordano acknowledges the support of the Max Planck Institute for Astronomy, Heidelberg, where part of this work has been done. We thank Philippe Jetzer, Asghar Qadir and Alexander F. Zakharov for the valuable discussions on the paper subject during many~years.}


\authorcontributions{\textbf{Author Contributions:} All authors have equally contributed to this paper and have read and approved the final~version.}


\conflictofinterests{\textbf{Conflicts of Interest:} The authors declare no conflict of interest.}


\bibliographystyle{mdpi}
\renewcommand\bibname{References}
\bibliographystyle{mdpi}
\makeatletter
\renewcommand\@biblabel[1]{#1. }
\makeatother


\begin{thebibliography}{999} 
\bibitem{einstein1911}
 Einstein, A. On the Influence of Gravitation on the
Propagation of Light. {\em Ann. Phys.} {\bf 1911}, {\em 340}, 898--908.

 \bibitem{dyson1919}
 Dyson, F.W.; Eddington, A.; Davidson, C. A determination
of the deflection of light by the sun's gravitational
field from observations made at the total eclipse of May
29, 1919. {\em Phil. Trans. Roy. Soc. A} {\bf 1920}, {\em 220}, 291--333.

 \bibitem{chwolson1924}
 Chwolson, O. \"Uber eine m\"ogliche Form fiktiver
Doppelsterne. {\em Astron. Nachr.} {\bf 1924}, {\em 221}, 329--330.

\bibitem{einstein1936}
Einstein, A. Lens-Like Action of a Star by the Deviation
of Light in the Gravitational Field. {\em Science} {\bf 1936}, {\em 84}, 506--507

\bibitem{zwicky1937a}
 Zwicky, F. Nebulae as Gravitational Lenses. {\em Phys.
Rev.} {\bf 1937}, {\em 51}, 290--290.

\bibitem{zwicky1937b}
 Zwicky, F. On the Probability of Detecting Nebulae which Act as Gravitational lenses. {\em Phys.
Rev.} {\bf 1937}, {\em 51}, 679--679.

 \bibitem{qso0957+561}
Walsh, D.; Carswell, R.F.; Weymann, R.J.
0957 + 561 A,B: Twin quasistellar objects or gravitational lens?
{\em Nature} {\bf 1979}, {\em 279}, 381--384.

 \bibitem{pacz1986}
 Paczy\`nski, B. Gravitational microlensing by the galactic halo. {\em Astrophys. J.} {\bf 1986}, {\em 304}, 1--5.

 \bibitem{johannsen}
 Johannsen, T. Sgr A$^*$ and general relativity. {\bf 2015}, {arXiv:1512.03818}.

 \bibitem{gwdetection}
Abbott, B.P.; Abbott, R.; Abbott, T.D.;   Abernathy, M.R.;  Acernese, F.;   Ackley, K.;  Adams, C.;   Adams, T.;  Addesso, P.;   Adhikari, R.X.;  {\it et al}. Observation of Gravitational Waves from a Binary Black Hole Merger. {\em Phys. Rev. Lett.} {\bf 2016}, {\em 116},  061102.

 \bibitem{SchneiderEhlersFalco}
 Schneider, P.; Ehlers, G.; Falco, E.E. Gravitational Lenses.  Springer Verlag: Berlin, Germany, {1992}.

 \bibitem{schmidt}
 Schmidt, M. 3C 273: A star-like object with large red-shift. {\em Nature} {\bf 1963}, {\em 197}, 1040--1040.

 \bibitem{pelt}
 Pelt, J.; Kayser, R.; Refsdal, S.; Schramm, T. The light curve and the time delay of QSO 0957+561. {\em Astron.~Astrophys.} {\bf 1996}, {\em 306}, 97--106.

 \bibitem{chen}
 Chen, G.H.; Kochanek, C.S.; Hewitt, J.N. The Mass Distribution of the Lens Galaxy in MG 1131+0456. {\em Astrophys. J.} {\bf 1995}, {\em 447}, 62--81.

 \bibitem{hennawi2008}
 Hennawi, J.F.; Gladders, M.D.; Oguri, M.; Dalal, N.; Koester, B.; Natarajan, P.; Strauss, M.A.; Inada, N.; Kayo, I.; Lin, H. A New Survey for Giant Arcs. {\em Astron. J.} {\bf 2008}, {\em 135}, 664--681.

 \bibitem{browne2003}
 Browne, I.W.A.; Wilkinson, P.N.; Jackson, N.J.F.; Myers, S.T.; Fassnacht, C.D.; Koopmans, L.V.E.; Marlow,~D.R.; Norbury, M.; Rusin, D.; Sykes, C.M.; \emph {et al}. The Cosmic Lens All-Sky Survey - II. Gravitational lens candidate selection and follow-up. {\em Mon. Not. R. Astron. Soc.} {\bf 2003}, {\em 341}, 13--32.

\bibitem{bolton2006}
Bolton, A.S.; Burles, S.; Koopmans, L.V.E.; Treu, T.; Moustakas, L.A. The Sloan Lens ACS Survey. I. A Large Spectroscopically Selected Sample of Massive Early-Type Lens Galaxies. {\em Astrophys. J.} {\bf 2006}, {\em 638}, 703--724.

 \bibitem{sdss}
Alam, S.; Albareti, F.D.; Allende, P.;  Anders, F.; Anderson, S.F.; Anderton, T.; Andrews, B.H.; Armengaud, E.; Aubourg, E.; Bailey, S.; \textit{et al}. The Eleventh and Twelfth Data Releases of the Sloan Digital Sky Survey: Final Data from SDSS-III. {\em Astrophys. J. Suppl. Ser.} {\bf 2015}, {\em 219}, 12.

 \bibitem{oguri2006}
 Oguri, M.; Inada, N.; Pindor, B.; Strauss, M.A.; Richards, G.T.; Hennawi, J.F.; Turner, E.L.; Lupton, R.H.; Schneider, D.P.; Fukugita, M. The Sloan Digital Sky Survey Quasar Lens Search. I. Candidate Selection Algorithm. {\em Astrophys. J.} {\bf 2006}, {\em 132}, 999--1013.

 \bibitem{blandford1992}
\scalebox{.98}[1.0]{ Blandford, R.D.; Narajan, R. Cosmological applications of gravitational lensing. {\em Annu. Rev. Astron. Astrophys.} } {\bf 1992}, {\em 30}, 311--358.

 \bibitem{treu2010}
 Treu, T. Strong lensing by galaxies. {\em Annu. Rev. Astron. Astrophys.} {\bf 2010}, {\em 48}, 87--125.

 \bibitem{kneib1996}
Kneib, J.-P.; Ellis, R.S.; Smail, I.; Couch, W. J.; Sharples, R.M. Hubble Space Telescope Observations of the Lensing Cluster Abell 2218. {\em Astrophys. J.} {\bf 1996}, {\em 471}, 643--656.

\bibitem{mao1998}
Mao, S.; Schneider, P. Evidence for substructure in lens galaxies? {\em Mon. Not. R. Astron. Soc.} {\bf 1998}, {\em 95}, 587--594.

\bibitem{metcalf2001}
Metcalf, R.B.; Madau, P. Compound Gravitational Lensing as a Probe of Dark Matter Substructure within Galaxy Halos. {\em Astrophys. J.} {\bf 2001}, {\em 563}, 9--20.

 \bibitem{xu2009}
 Xu, D.D.; Mao, S.; Wang, J.;  Springel, V.; Gao, L.; White, S.D.M.; Frenk, C.S.; Jenkins, A.; Li, G.; Navarro, J.F. Effects of dark matter substructures on gravitational lensing: results from the Aquarius simulations. {\em Mon. Not. R. Astron. Soc.} {\bf 2009}, {\em 408}, 1235--1253.

\bibitem{metcalf2012}
 Metcalf, R. B.; Amara, A. Small-scale structures of dark matter and flux anomalies in quasar gravitational lenses. {\em Mon. Not. R. Astron. Soc.} {\bf 2012}, {\em 419}, 3414--3425.

 \bibitem{xu2015}
Xu, D.D.; Sluse, D.; Gao, L.  Wang, J.; Frenk, C.; Mao, S.; Schneider, P.; Springel, V. How well can cold dark matter substructures account for the observed radio flux-ratio anomalies. {\em Mon. Not. R. Astron. Soc.} {\bf 2015}, {\em 447}, 3189--3206.

 \bibitem{vegetti2012}
 Vegetti, S.; Lagattuta, D.J.; McKean, J.P.; Auger, M. W.; Fassnacht, C.D.; Koopmans, L.V.E. Gravitational detection of a low-mass dark satellite galaxy at cosmological distance. {\em Nature} {\bf 2012}, {\em 481}, 341--343.

 \bibitem{vegetti2014}
 Vegetti, S.; Koopmans, L.V.E. Auger, M.W. Inference of the cold dark matter substructure mass function at z = 0.2 using strong gravitational lenses. {\em Mon. Not. R. Astron. Soc.} {\bf 2014},
 {\em 442}, 2017--2035.

 \bibitem{kochanek2004}
 Kochanek, C.S. Quantitative Interpretation of Quasar Microlensing Light Curves. {\em Astrophys. J.} {\bf 2004}, {\em 605}, 58--77.

 \bibitem{mosquera2011}
 Mosquera, A. M., Kochanek, C.S. The Microlensing Properties of a Sample of 87 Lensed Quasars. \mbox{{\em Astrophys. J.}} {\bf 2011}, {\em 738},  96.

\bibitem{pooley2012}
Pooley, D.; Rappaport, S.; Blackburne, J.A.; Schechter, P.L.; Wambsganss, J. X-Ray and Optical Flux Ratio Anomalies in Quadruply Lensed Quasars. II. Mapping the Dark Matter Content in Elliptical Galaxies. {\em Astrophys. J.} {\bf 2012}, {\em 744},  111.

 \bibitem{soucail1987}
 Soucail, G.; Fort, B.; Mellier, Y.; Picat, J.P. A blue ring-like structure, in the center of the A 370 cluster of galaxies. {\em Astron. Astrophys.} {\bf 1987}, {\em 172}, L14--L16.

 \bibitem{stark2008}
 Stark, D.P.; Swinbank, A.M.; Ellis, R.S.  Dye, S.; Smail, I.R; Richard, J. The formation and assembly of a typical star-forming galaxy at redshift z$\sim$3. {\em Nature} {\bf 2008}, {\em 455}, 775--777.

 \bibitem{kelly2015}
 Kelly, P.L.; Rodney, S.A.; Treu, T.; Foley, R.J.; Brammer, G.; Schmidt, K.B.; Zitrin, A.; Sonnenfeld, A.; \mbox{Strolger, L.-G.;} Graur, O.; \emph {et al}. Multiple images of a highly magnified supernova formed by an early-type cluster galaxy lens. {\em Science} {\bf 2015}, {\em 347}, 1123--1126.

 \bibitem{treu2016}
 Treu, T.; Brammer, G.; Diego, J.M.; Grillo, C.; Kelly, P.L.; Oguri, M.; Rodney, A.; Rosati, P.; Sharon, K.; Zitrin, A.``Refsdal'' Meets Popper: Comparing Predictions of the Re-appearance of the Multiply Imaged Supernova Behind MACSJ1149.5+2223. {\em Astrophys. J.} {\bf 2016}, {\em 817}, 60.

 \bibitem{refsdal1964}
 Refsdal, S. On the possibility of determining Hubble's parameter and the masses of galaxies from the gravitational lens effect.
 {\em Mon. Not. R. Astron. Soc.} {\bf 1964}, {\em 128}, 307--310.

 \bibitem{jackson2007}
 Jackson, N. The Hubble Constant. {\em Living Rev. Relativ.} {\bf 2007}, {\em 10},  4.

 \bibitem{suyu2010}
 Suyu, S.H.; Marshall, P.J.; Auger, M.W.; Hilbert, S.; Blandford, R.D.; Koopmans, L.V.E.; Fassnacht, C.D.; Treu, T.; \textit{et al}. Dissecting the Gravitational lens B1608+656. II. Precision Measurements of the Hubble Constant, Spatial Curvature, and the Dark Energy Equation of State. {\em Astrophys. J.} {\bf 2010}, {\em 711}, 201--221.

 \bibitem{cosmograil2016}
 Bonvin, V.; Tewes, M.; Courbin, F.; 	
	 Kuntzer, T.; Sluse, D.; Meylan, G. COSMOGRAIL: the COSmological MOnitoring of GRAvItational Lenses. XV. Assessing the achievability and precision of time-delay measurements. {\em Astron. Astrophys.} {\bf 2016}, {\em 585} id.A88.

 \bibitem{bartelmann2001}
 Bartelmann, M.; Schneider, P. Weak gravitational lensing. {\em Phys. Rep.} {\bf 2001}, {\em 340}, 291--472.


 \bibitem{tyson1990}
 Tyson, J.A.; Wenk, R.A.; Valdes, F. Detection of systematic gravitational lens galaxy image alignments---Mapping dark matter in galaxy clusters. {\em Astrophys. J.} {\bf 1990}, {\em 349}, L1--L4.

\bibitem{bacon2000}
Bacon, D.J.; Refregier, A.R.; Ellis, R.S. Detection of weak gravitational lensing by large-scale structure. {\em Mon.~Not. R. Astron. Soc.} {\bf 2000}, {\em 318}, 625--640.

\bibitem{kaiser2000}
Kaiser, N. A New Shear Estimator for Weak-Lensing Observations. {\em Astrophys. J.} {\bf 2000}, {\em 537}, 555--577.

 \bibitem{seitz1996}
 Seitz, S.; Schneider, P. Cluster lens reconstruction using only observed local data: An improved finite-field inversion technique. {\em Astron. Astrophys.} {\bf 1996}, {\em 305}, 383--401.

\bibitem{lombardi1999}
\scalebox{.96}[1.0]{Lombardi, M.; Bertin, G. A fast direct method of mass reconstruction for gravitational lenses. {\em Astron. Astrophys.}} {\bf 1999}, {\em 348}, 38--42.

 \bibitem{schneider2000}
 Schneider, P.; King, L.; Erben, T. Cluster mass profiles from weak lensing: Constraints from shear and magnification information. {\em Astron. Astrophys.} {\bf 2000}, {\em 353}, 41--56.

 \bibitem{han2015}
Han, J.; Eke, V.R.; Frenk, C.S.; Mandelbaum, R.; Norberg, P.; Schneider, M.D.; Peacock, J.A.; Jing, Y.; Baldry, I.; Bland-Hawthorn, J.; \emph {et al}. Galaxy And Mass Assembly (GAMA): the halo mass of galaxy groups from maximum-likelihood weak lensing. {\em Mon. Not. R. Astron. Soc.} {\bf 2015}, {\em 446}, 1356--1379.

\bibitem{marshall2002}
Marshall, P.J.; Hobson, M.P.; Gull, S.F.; Bridle, S.L. Maximum-entropy weak lens reconstruction: improved methods and application to data. {\em Mon. Not. R. Astron. Soc.} {\bf 2002}, {\em 335}, 1037--1048.

 \bibitem{clowe2004}
 Clowe, D.; De Lucia, G.; King, L. Effects of asphericity and substructure on the determination of cluster mass with weak gravitational lensing. {\em Mon. Not. R. Astron. Soc.} {\bf 2004}, {\em 350}, 1038--1048.

 \bibitem{corless2007}
 Corless, V.L.; King, L.J. A statistical study of weak lensing by triaxial dark matter haloes: consequences for parameter estimation. {\em Mon. Not. R. Astron. Soc.} {\bf 2007}, {\em 380}, 149--161.

 \bibitem{hoekstra2007}
 Hoekstra, H. A comparison of weak-lensing masses and X-ray properties of galaxy clusters. {\em Mon. Not. R. Astron. Soc.} {\bf 2007}, {\em 379}, 317--330.

\bibitem{zhang2008}
 Zhang, Y.-Y.; Finoguenov, A.; B\"ohringer, H.; Kneib, J.-P.; Smith, G.P.; Kneissl, R.; Okabe, N.; Dahle, H. LoCuSS: Comparison of observed X-ray and lensing galaxy cluster scaling relations with simulations. {\em Astron. Astrophys.} {\bf 2008}, {\em 482}, 451--472.

 \bibitem{contaldi2003}
 Contaldi, C.R.; Hoekstra, H.; Lewis, A. Joint Cosmic Microwave Background and Weak Lensing Analysis: Constraints on Cosmological Parameters. {\em Phys. Rev. Lett.} {\bf 2003}, {\em 22},  221303.

 \bibitem{hollenstein2009}
 Hollenstein, L.; Sapone, D.; Crittenden, R.; Sch\"afer, B.M. Constraints on early dark energy from CMB lensing and weak lensing tomography. {\em J. Cosmol. Astropart. Physucs} {\bf 2009}, {\em
 04}, id. 012.

 \bibitem{majerotto2016}
 Majerotto, E.; Sapone, D.; Sch\"afer, B,M. Combined constraints on deviations of dark energy from an ideal fluid from Euclid and Planck. {\em Mon. Not. R. Astron. Soc.} {\bf 2016}, {\em 456}, 109--118.

 \bibitem{clowe2006}
Clowe, D.; Brada\^c, M.; Gonzalez, A.H.; Markevitch, M.   Markevitch, M.;  Randall, S.W.;  Jones, C.;   Zaritsky, D. A Direct Empirical Proof of the Existence of Dark Matte. {\em Astrophys. J.} {\bf 2006}, {\em 648}, L109--L113.

 \bibitem{amendola} 	
 Amendola, L.; Appleby, S.; Bacon, D.  Baker, T.   Baldi, M.
 Bartolo, N.  Blanchard, A.  Bonvin, C.  Borgani, S. 
Branchini, E.; \textit{et al}. Cosmology and Fundamental Physics with the Euclid Satellite. {\em Living Rev. Relat.} {\bf 2013}, {\em 16}, 6.

 \bibitem{chang2013}
 Chang, C.; Jarvis, M.; Jain, B.       Kahn, S.M.;  Kirkby, D.;  Connolly, A.;  Krughoff, S.;  Peng, E.-H.; Peterson, J.R. The effective number density of galaxies for weak lensing measurements in the LSST project. {\em Mon. Not. R. Astron. Soc.} {\bf 2013}, {\em 434}, 2121--2135.

 \bibitem{Heymans2003}
 Heymans, C.; Heavens, A. Weak gravitational lensing: reducing the contamination by intrinsic alignments. {\em Mon. Not. R. Astron. Soc.} {\bf 2003}, {\em 339}, 711--720.

 \bibitem{Valageas2014}
\scalebox{.97}[1.0]{Valageas, P. Source-lens clustering and intrinsic-alignment bias of weak-lensing estimators. {\em Astron. Astrophys.}} {\bf 2014}, {\em 561},  A53.

 \bibitem{mao2012}
 Mao, S. Astrophysical applications of gravitational microlensing. {\em Res. Astron. Astrophys.} {\bf 2012}, {\em 12}, 947--972.

 \bibitem{Perryman14}
Perryman, M. \emph {The Exoplanet Handbook}; {University Press}: Cambridge, UK, 2014.

\bibitem{sumi}
Sumi, T.; Kamiya, K.; Udalski, A.; Bennett, D.P.; Bond, I.A.; Abe, F.; Botzler, C.S.; Fukui, A.; Furusawa, K.; Hearnshaw, J.B.; Itow, Y.; \emph {et al}. Unbound or distant planetary mass population detected by gravitational microlensing. {\em Nature} {\bf 2011}, {\em 473}, 349--352.

 \bibitem{Ingrosso2009}
Ingrosso, G.; Calchi Novati, S.; De Paolis, F.; Jetzer, P.; Nucita, A.A.; Strafella, F. Pixel lensing as a way to detect extrasolar planets in M31. {\em Mon. Not. R. Astron. Soc.} {\bf 2009}, {\em 399}, 219--228.

\bibitem{an} An, J.H.; Evans, N.W.; Kerins, E.; Baillon, P.; Calchi-Novati, S.; Carr, B.J.; Creze, M.; Giraud-Heraud, Y.; Gould, A.; Hewett, P.; \emph {et al}. The Anomaly in the Candidate Microlensing Event PA-99-N2. {\em Astrophys. J.} {\bf 2004}, {\em 601}, 845--857.

\bibitem{dominik2000}
Dominik, M.; Sahu, K.C. Astrometric Microlensing of Stars.  {\em Astrophys. J.} {\bf 2000}, {\em 534}, 213--226.

\bibitem{lee2010}
Lee, C.-H.; Seitz, S.; Riffeser, A.; Bender, R. Finite-source and finite-lens effects in astrometric microlensing. {\em Mon. Not. R. Astron. Soc.} {\bf 2010}, {\em 407}, 1597--1608.

\bibitem{walker1995}
Walker, M.A. Microlensed Image Motions. {\em Astrophys. J.} {\bf 1995}, {\em 453}, 37--39.

\bibitem{eyer}
Eyer, L.; Holl, B.; Pourbaix, D.;   Mowlavi, N.;  Siopis, C.;  Barblan, F.;  Evans, D.W.;  North, P. The Gaia Mission. {\em Cent. Eur. Astrophys. Bull. (CEAB)} {\bf 2013}, {\em 37}, 115--126.

\bibitem{paczinsky1996}
Paczy\'{n}ski, B. The Masses of Nearby Dwarfs and Brown Dwarfs with the HST. {\em Acta Astron.} {\bf 1996}, {\em 46}, 291--296.

 \bibitem{proft2011}
 Proft, S.; Demleitner, M.; Wambsganss, J. Prediction of astrometric microlensing events during the Gaia mission. {\em Astron. Astrophys.} {\bf 2011}, {\em 536}, A50.

 \bibitem{pol1}
 Ingrosso, G.; Calchi Novati, S.; De Paolis, F.; Jetzer, P.; Nucita, A.A.; Strafella, F.; Zakharov, A.F. Polarization in microlensing events towards the Galactic bulge. {\em Mon. Not. R. Astron. Soc.} {\bf 2012}, {\em 426}, 1496--1506.

 \bibitem{chandra} Chandrasekhar, S. \emph {Radiative Transfer}; Clarendon Press: Oxford, UK, 1950.

\bibitem{simmons2002}
Simmons, J.F.L.; Bjorkman, J.E.; Ignace, R.; Coleman, I.J. Polarization from microlensing of spherical circumstellar envelopes by a point lens. {\em Mon. Not. R. Astron. Soc.} {\bf 2002}, {\em 336}, 501--510.

 \bibitem{ingrossopol2015} Ingrosso, G.; Calchi Novati, S.; De Paolis, F.; Jetzer, P.; Nucita, A.A.; Strafella, F. Measuring polarization in microlensing events. {\em Mon. Not. R. Astron. Soc.} {\bf 2015}, {\em 446}, 1090--1097.

 \bibitem{Henderson2014}
 Henderson, C.B.; Gaudi, B.S.; Han, C.; Skowron, J.; Penny, M.T.; Nataf, D.; Gould, A.P. Optimal Survey Strategies and Predicted Planet Yields for the Korean Microlensing Telescope Network. {\em Astrophys. J.} {\bf 2014}, {\em 794},  52.

 \bibitem{Penny2013}
 Penny, M.T.; Kerins, E.; Rattenbury, N.; Beaulieu, J.-P.; Robin, A.C.; Mao, S.; Batista, V.; Calchi Novati, S.; Cassan, A.; Fouque, P.; \emph {et al}. Zapatero Osorio ExELS: An exoplanet legacy science proposal for the ESA Euclid mission - I. Cold exoplanets. {\em Mon. Not. R. Astron. Soc.} {\bf 2013}, {\em 434}, 2--22.

 \bibitem{Yee2014}
 Yee, J. C.; Albrow, M.; Barry, R.K.; Bennett, D.; Bryden, G.; Chung, S.-J.; Gaudi, B.S.; Gehrels, N.; \mbox{Gould, A.; } Penny, M.T.; \emph {et al}. Takahiro SumiNASA ExoPAG Study Analysis Group 11: Preparing for the WFIRST Microlensing Survey. {\bf 2014}, {arXiv:1409.2759}.

 \bibitem{2000ApJ...529...69H}
 Heyrovsk\'y, D.; Sasselov, D. Detecting Stellar Spots by Gravitational Microlensing. \emph{Astrophys. J.} {\bf 2000}, {\em 529}, 69--76.

 \bibitem{2002MNRAS.335..539H} Hendry, M.A.; Bryce, H.M.;
  Valls-Gabaud, D. The microlensing signatures of photospheric starspots. \mbox{\emph{Mon. Not. R. Astron.
  Soc.}} {\bf 2002}, {\em 335}, 539--549.

 \bibitem[Giordano et al.(2015)]{2015MNRAS.453.2017G} Giordano, M.; Nucita,
 A.A.; De Paolis, F.; Ingrosso, G. Starspot induced effects in microlensing events with rotating source star. \emph{Mon. Not.
  R. Astron. Soc.} {\bf 2015}, {\em 453}, 2017--2021.

 \bibitem[Kim et al.(2010)]{2010SPIE.7733E..3FK} Kim, S.-L.; Park, B.-G.; Lee,
 C.-U.; Yuk, I.-S.; Han, C.; O'Brien, T.; Gould, A.; Lee, J.W.; Kimet, D.-J. Technical specifications of the KMTNet observation system. \emph{Proc.~ SPIE} {\bf 2010}, doi:10.1117/12.856833.

 \bibitem[Sajadian(2015)]{2015MNRAS.452.2587S} Sajadian, S. Detecting stellar spots through polarimetric observations of microlensing events in caustic-crossing. \emph{Mon.
  Not. R. Astron. Soc.} {\bf 2015}, {\em 452}, 2587--2596.

 \bibitem[Dominik(1999)]{1999A&A...349..108D} Dominik, M. The binary gravitational lens and its extreme cases. \emph{Astron.
  Astrophys.} {\bf 1999}, {\em 349}, 108--125.

 \bibitem[An(2005)]{2005MNRAS.356.1409A} An, J.~H. Gravitational lens under perturbations: symmetry of perturbing potentials with invariant caustics. \emph{Mon. Not.
 R. Astron. Soc.} {\bf 2005}, {\em 356}, 1409--1428.

 \bibitem[Griest \& Safizadeh(1998)]{1998ApJ...500...37G} Griest, K.;
 Safizadeh, N. The Use of High-Magnification Microlensing Events in Discovering Extrasolar Planets. \emph{Astrophys. J.} {\bf 1998}, {\em 500}, 37--50.

 \bibitem[Penny et al.(2011)]{2011MNRAS.417.2216P} Penny, M.~T.; Kerins, E.;
 Mao, S. Rapidly rotating lenses: Repeating features in the light curves of short-period binary microlenses. \emph{Mon. Not. R. Astron. Soc.} {\bf 2011}, {\em 417},
 2216--2229.

 \bibitem[Nucita et al.(2014)]{2014MNRAS.438.2466N} Nucita, A.~A.; Giordano, M.;
 De Paolis, F.; Ingrosso, G. Signatures of rotating binaries in microlensing experiments. \emph{Mon. Not. R.
  Astron. Soc.} {\bf 2014}, {\em 438}, 2466--2473.

 \bibitem{chandraBH}
 Chandrasekhar, S. \emph {The Mathematical Theory of Black Holes}; {Oxford University Press}: Oxford, UK, 1983.

 \bibitem{holzwheeler} Holz, D.E.; Wheeler, J.A. Retro-MACHOs: pi in the Sky. {\em Astrophys. J.} {\bf 2002}, {\em 57}, 330--334.

 \bibitem{depaolisS2} De Paolis, F.; Geralico, A.; Ingrosso, G.; Nucita, A.A. The black hole at the galactic center as a possible
retro-lens for the S2 orbiting star. {\em Astron. Astrophys.} {\bf 2003}, {\em 409}, 809--812.

\bibitem{gillessen}
Gillessen, S.; Eisenhauer, F.; Trippe, S.; Alexander, T.; Genzel, R.; Martins, F.; Ott, T. Monitoring Stellar Orbits Around the Massive Black Hole in the Galactic Center. {\em Astrophys. J.} {\bf 2009}, {\em 692}, 1075--1109.

\bibitem{doku2014}
Dokuchaev, V.I. Spin and mass of the nearest supermassive black hole. {\em Gen. Rel. Gravit.} {\bf 2014}, {\em 46},  1832.

\bibitem{doku2015}
Dokuchaev, V.I.; Eroshenko, Y.N. Physical Laboratory at the center of the Galaxy. {\em Phys. Uspekhi} {\bf 2015}, {\em 58}, 772--784.

\bibitem{iorio2012}
Iorio, L. Perturbed stellar motions around the rotating black hole in Sgr A* for a generic orientation of its spin axis. {\em Phys. Rev. D} {\bf 2012}, {\em 84},  124001.

\bibitem{zak2007}
Zakharov, A.F.; Nucita, A.A.; De Paolis, F.; Ingrosso, G. Apoastron shift constraints on dark matter distribution at the Galactic Center. {\em Phys. Rev. D} {\bf 2007}, {\em 76}, 062001.

\bibitem{zak2012}
Zakharov, A.F.; de Paolis, F.; Ingrosso, G.; Nucita, A.A. Shadows as a tool to evaluate the black hole parameters and a dimension of spacetime. {\em New Astron. Rev.} {\bf 2012}, {\em 56}, 64--73.

\bibitem{falcke2013}
Falcke, H.; Markoff, S.B. Toward the event horizon---The supermassive black hole in the galactic center. {\em Classicala Quantum Gravit.} {\bf 2013}, {\em 30}, 244003.

\bibitem{zak2014}
Zakharov, A.F.; Borka, D.; Borka Jovanovi\'c, V.; Jovanovi\'c, P. Constraints on $R^n$ gravity from precession of orbits of S2-like stars: A case of a bulk distribution of mass. {\em Adv. Space Res.} {\bf 2014}, {\em 54}, 1108--1112.

\bibitem{zak}
Zakharov, A.F.; Nucita, A.A.; De Paolis, F.; Ingrosso, G. Solar system constraints on $R^n$ gravity. \mbox{{\em Phys. Rev. D}} {\bf 2006}, {\em 74},  107101.

 \bibitem{pasp2007}
 Nucita, A.A.; De Paolis, F.; Ingrosso, G.; Qadir, A.; Zakharov, A.F. Sgr A*: A Laboratory to Measure the Central Black Hole and Stellar Cluster Parameters. {\em Publ. Astron. Soc. Pac.} {\bf 2007}, {\em 119}, 349--359.

\bibitem{plewa2015}
Plewa, P.M. et al. Pinpointing the near-infrared location of Sgr A$^*$ by correcting optical distortion in the NACO imager {\bf 2015}, arXiv: 1509.01941.

\bibitem{depaolis2011} De Paolis, F.; Ingrosso, G.; Nucita, A.A.; Qadir, A.; Zakharov, A.F. Estimating the parameters of the Sgr A$^*$ black hole. {\em Gen. Rel. Gravit.} {\bf 2011}, {\em 43}, 977--988.

\bibitem{zak2005} Zakharov, A.F.; De Paolis, F.; Ingrosso, G.; Nucita, A. Direct measurements of black hole charge with future astrometrical missions. {\em Astron. Astrophys.} {\bf 2005}, {\em 442}, 795--799.

\bibitem{bozzamancini2004}
Bozza, V.; Mancini, L. Gravitational Lensing by Black Holes: A Comprehensive Treatment and the Case of the Star S2. {\em Astrophys. J.} {\bf 2005}, {\em 611}, 1045--1053.

\bibitem{doeleman} Doeleman, S.S.; Weintroub, J.; Rogers, A.E.E.; Plambeck, R.; Freund, R.; Tilanus, R.P.J.; Friberg, P.; Ziurys, L.M.; Moran, J.M.; Corey, B.; \emph {et al}. Event-horizon-scale structure in the supermassive black hole candidate at the Galactic Centre. {\em Nature} {\bf 2008}, {\em 455}, 78--80.


\end{thebibliography}


%


%

\end{document}